\documentclass[onecolumn]{aa} 
\usepackage{lineno}           

\usepackage[pdftex=true,colorlinks=true,linkcolor=blue,citecolor=blue,urlcolor=black]{hyperref}
\usepackage{graphicx}
\usepackage{txfonts}
\usepackage{varwidth}

\usepackage{xspace}
\usepackage{xcolor}
\usepackage{ulem}
\newcommand{\red}{\textcolor{black}}
\newcommand{\magenta}{\textcolor{black}}
\newcommand{\fpc}[1]{\textcolor{black}{#1}}

\newcommand{\Fermi}{\textit{Fermi}\xspace}
\newcommand{\band}{\texttt{Band}\xspace}
\newcommand{\issm}{\texttt{ISSM}\xspace}
\newcommand{\logp}{\texttt{LP}\xspace}
\newcommand{\logpm}{\texttt{LP1}\xspace}
\newcommand{\logpmm}{\texttt{LP2}\xspace}
\newcommand{\cutpl}{\texttt{CUTPL}\xspace}
\newcommand{\cutbpl}{\texttt{CUTBPL}\xspace}
\newcommand{\cstat}{$C_\mathrm{stat}$\xspace}
\newcommand{\pgstat}{$PG_\mathrm{stat}$\xspace}
\usepackage{amsmath}
\usepackage{lipsum}
\usepackage{multirow}
\newcommand{\GRBBone}{\mathrm{GRB}$\_$\mathrm{B001}\xspace}
\newcommand{\GRBBten}{\mathrm{GRB}$\_$\mathrm{B010}\xspace}
\newcommand{\GRBBhun}{\mathrm{GRB}$\_$\mathrm{B100}\xspace}
\newcommand{\unit}{keV$^{-1}$\,cm$^{-2}$\,s$^{-1}$\xspace}
\def\ergfluence{erg\,cm$^{-2}$\xspace}
\usepackage{lineno}
\begin{document} 
\title{A new fitting function for GRB MeV spectra based on the internal shock synchrotron model}
\author{
  M.~Yassine\inst{1,2}
  \and
  F.~Piron\inst{3}
  \and
  F.~Daigne\inst{4}
  \and
  R.~Mochkovitch\inst{4}
  \and
  F.~Longo\inst{1,2}
  \and
  N.~Omodei\inst{5}
  \and
  G.~Vianello\inst{5}
}

\institute{
    Istituto Nazionale di Fisica Nucleare, Sezione di Trieste, I-34127 Trieste, Italy
  \and
    Dipartimento di Fisica, Università di Trieste, I-34127 Trieste, Italy \\
  \email{mychbib@gmail.com, francesco.longo@ts.infn.it}
  \and
    Laboratoire Univers et Particules de Montpellier, Université de Montpellier, CNRS/IN2P3, Montpellier, France\\
  \email{piron@in2p3.fr}
  \and
  UPMC-CNRS, UMR7095, Institut d’Astrophysique de Paris, 75014 Paris, France\\
  \email{daigne@iap.fr, mochko@iap.fr}
  \and
  W. W. Hansen Experimental Physics Laboratory,  Kavli Institute for Particle Astrophysics and Cosmology,  Department of Physics and
SLAC National Accelerator Laboratory, Stanford University, Stanford, CA 94305, USA\\
}

\date{}

\abstract  
  {}
  {The physical origin of the GRB prompt emission is still a subject of debate.
\fpc{Internal shock models have been widely explored owing to their ability to explain most of the high-energy
properties of this emission phase. 
While the \band function or other phenomenological functions are commonly used to fit GRB prompt emission spectra, we propose a new
parametric function that is inspired by an internal shock physical model. We use this function as a proxy of the model to confront it easily to GRB observations.
}
    }
    {
\fpc{We built a parametric function that represents the spectral form of the synthetic bursts provided by our internal
shock synchrotron model (\issm).
We simulated the response of the \Fermi instruments to the synthetic bursts and fitted the obtained count spectra to
validate the \issm function.
Then, we applied this function to a sample of 74 bright GRBs detected by the \Fermi/GBM, and we computed the width of
their spectral energy distributions around their peak energy.
For comparison, we fitted also the phenomenological functions that are commonly used in the literature.
Finally, we performed a time-resolved analysis of the broadband spectrum of GRB\,090926A, which was jointly detected by the \Fermi GBM and
LAT.
 This spectrum has a complex shape and exhibits a power-law component with an exponential cutoff at high energy, which is
 compatible with inverse Compton emission attenuated by gamma-ray internal absorption.
}
}
   {This work proposes a new parametric function for spectral fitting that is based on a physical model.
   The \issm function reproduces $81$\% of the \fpc{spectra in the GBM bright} GRB sample, versus $59$\% \fpc{for} the \band
   function, for the same number of parameters.
\fpc{It gives also relatively good fits to the GRB\,090926A spectra.
The width of the MeV spectral component that is obtained from the fits of the \issm function is slightly larger than the
width from the \band fits, but it is smaller when observed over a wider energy range.
Moreover, all of the 74 analysed spectra are found to be significantly wider than the synthetic synchrotron spectra.
We discuss possible solutions to reconcile the observations with the internal shock synchrotron model, such as an improved modeling
of the shock micro-physics or more accurate spectral measurements at MeV energies.} 
 } 
   {}

\maketitle

\section{Introduction}
\label{sec:intro}
\fpc{Gamma-ray bursts (GRBs) were discovered more than fifty years ago, and they are the most electro-magnetic events ever observed in the Universe.}
They are brief flashes of high-energy radiation \fpc{emitted by an ultra-relativistic collimated outflow which is thought
  to originate from a stellar-mass black hole formed by the merging of binary systems \citep{Nakar2007PhysRep442-166, DAvanzo2015JHEAp7-73} or the explosions of
  massive stars} \citep{Woosley2006ARAA44-507, stanek, kawabata, hjorth1, bloom, hjorth2, gehrels, abbott}. 
\fpc{GRB emission is observed} in two \fpc{successive} phases, a
short \fpc{phase of} intense \fpc{radiation} followed by a long-lived afterglow phase.
\fpc{While both emissions are essentially non thermal, the}
prompt phase is \fpc{notably} characterized by the irregular shape and the fast variability of its temporal profile.
\fpc{Despite substantial efforts in modeling the GRB prompt emission, different scenarios such as internal shocks \citep{rees}, dissipative
  photospheres \citep{BM17} or reconnection above the photosphere \citep{giannios, mckinney, sironi,BG16} have been proposed to explain its physical origin.
Internal shock models have been explored in detail \citep{kobayashi, DM98, BD09, DB11, BD14} owing to their ability to produce
emissions from the visible to the GeV domain and to account for GRB observed properties such as their spectral evolution
and the} extreme variability seen in their light curves.
\fpc{In this class of models,} the GRB 
relativistic outflow 
converts a fraction of its kinetic energy into internal energy through internal shocks,
\fpc{which occur} when the distribution of the Lorentz factors in the flow is highly non-uniform.
\fpc{Part} of the energy that is dissipated in \fpc{the} shocks is transferred to a fraction of the electrons which emit
non-thermal synchrotron and inverse Compton radiations.\\


Since the launch of the \Fermi satellite in \fpc{June} 2008, the GRB
\fpc{high-energy emission has been studied with great sensitivity.
The Large Area Telescope}
(LAT, 20 MeV- 300 GeV, \citep{Atwood2009ApJ6971071A}) has detected 
more than 180 GRBs \citep{LatGrbCatalog2}
\fpc{ thanks to its}
wide field of view ($2.4$ sr), its large effective area ($\sim$ 0.9 m$^2$ above $\sim$1 GeV) and to the
\fpc{improved event reconstruction (Pass 8 hereafter) that has been implemented in 2015 \citep{Atwood2013}.}
\fpc{The \fpc{Gamma-ray Burst Monitor} (GBM) is the second instrument \fpc{onboard \Fermi} and it consists of 12 sodium
  iodide (NaI, 8 keV - 1 MeV) and 2 bismuth germanate (BGO, 250 keV- 40 MeV) detectors placed around the \Fermi spacecraft.
  The GBM monitors continuously a large portion of the sky ($9.5$ sr), and it has detected more than 2600 GRBs so far \citep{2016ApJS22328N}.
Together, the GBM and the LAT cover more than seven decades in energy, hence they are the most suitable instruments currently in
operations to study the broadband high-energy emission of GRBs.\\
} 

The keV-MeV spectral component of GRBs, which is often attributed to synchrotron emission, is commonly fitted by the
phenomenological \band function \citep{Band1993}. Despite its ability to describe many of the GRB non-thermal
spectra, this function has little physical grounds and is not suitable for a fair fraction of spectra (see, e.g.,
\cite{Gruber2014}).
The interpretation of the GRB spectral fit results faces another problem that has been pointed out twenty years ago by
\cite{Preece1998} (see also \cite{Crider1997, Ghisellini2000, Burgess2015}). In their analysis of CGRO/BATSE bursts, these authors came to the conclusion that most of the fitted
spectral slopes are too hard to be compatible with the expectations from the synchrotron theory at low energy, an issue
that is now refered to as the ``synchrotron line-of-death problem'' \citep{Ghisellini2000,Axelsson2015,Burgess2015}. 

More recently, \cite{Yu2015A&A} and \cite{Axelsson2015} used the spectral sharpness to show that the spectrum that is expected from an
electron synchrotron model is wider than the \band spectra of most GRBs detected by the GBM, calling for a new physical interpretation of the keV-MeV spectral component.
However, it should be noted that the theoretical spectrum considered in \cite{Yu2015A&A} was essentially derived from a
pure Maxwellian electron distribution,
which does not account for the dynamical evolution of the electron and photon distributions in the GRB jet.
In addition, the authors did not attempt to fit this theoretical model to the data, which might introduce
instrumental biases in the comparison with the \band fit results. Actually, direct fits of the synchrotron emission model to GRB prompt spectra have been performed by \cite{zhang2016} and \cite{Burgess2019}, who showed that the line-of-death and spectral sharpness issues are likely artefacts due to the use of the \band function. In the same spirit, this work compares the predictions of an actual internal shock synchrotron model to the observations.
\\


In this work, we investigated the \fpc{version of the internal shock model described in \cite{DM98,BD09,DB11,BD14}}.
We simulated synthetic bursts provided by this model using the GBM and LAT detector responses.
The characteristics of the synthetic bursts and our simulation procedure are described in
Sect.~\ref{sec:sim}.
In Sect.~\ref{sec:models} we present the \fpc{functions} used to fit the 
burst spectra, \fpc{including a new fitting function (called \issm hereafter) that is directly built from the synthetic
  spectra in the keV-MeV energy range}.
\fpc{The spectral analysis of the synthetic bursts and the computation of their spectral width
are reported} in Sect.~\ref{sec:ana_sim}.
\fpc{
In Sect.~\ref{sec:ana_gbm}, we apply the same set of fitting functions to a sample of 74 GBM bright GRBs.
The data selection and the technique of identification of the best fit spectral model are presented, as well as a focus on
the spectral parameters and sharpness obtained for the \band and \issm functions. 
In Sect.~\ref{sec:ana_090926A} we revisit the spectral analysis of GRB\,090926A using the new \issm function. 
This burst was bright in the GBM and LAT instruments, and it exhibits fast variability above $100$ MeV during the keV-MeV prompt emission.
As reported in \citep{Yassine2017} (Y17 hereafter), it constitutes an ideal case to test the internal shock model from keV to GeV energies.
Finally, we discuss our results in Sect.~\ref{sec:discussion} and give our conclusions in Sect.~\ref{sec:concl}.
}

\section{Simulation of the synthetic bursts}
\label{sec:sim}
%

\subsection{The internal shock model}

The version of the internal shock model that we used is able to reproduce most of the GRB properties, in particular the
variability timescales and the shape of the GRB light curves \citep{DM98}.
In this model,
the GRB outflow consists of a set of solid layers which move at different Lorentz factors,
whose collisions mimick the propagation of an internal shock wave
along the GRB jet.
Each GRB is characterized by its redshift, duration and kinetic energy, and by a Lorentz factor profile.   
The model also assumes that some fraction $\epsilon_B$ of the energy dissipated in the shocks
is transferred to the magnetic field, and that a fair fraction $\epsilon_e$ is injected into a small part
$\zeta$ of accelerated electrons.
The energy distribution of the accelerated electrons is a power law with a slope $-p$ which 
is set to a value ranging from $-2.9$ to $-2.3$. This adopted interval for the index of the electron distribution corresponds to a typical high energy spectral index $\beta$=-($p$/2+1) between 2.15 and 2.45 as observed \citep{DM98}. 
In addition to the GRB outflow dynamics,
the model accounts for the main radiative processes at high energy. 
The numerical code that simulates the shock dynamics has been coupled to a radiative code, which follows the
evolution of the electron and photon distributions in order to produce realistic light curves and spectra from keV to GeV
energies in the observer frame \citep{BD09}.
The radiative processes include the synchrotron emission from the accelerated electrons and the inverse Compton (IC)
scatterings in the Thomson and Klein-Nishina regimes. Synchrotron self-absorption at low energy and photon-photon
annihilation at high energy are also accounted for.

\subsection{Characteristics of the bursts}


The synthetic burst that we considered corresponds to the case B of \citep{BD14} (BD14 hereafter) owing to its typical kinetic energy,
$E_k$=10$^{54}$ erg, and to its brightness in the LAT energy range.
The burst is long, with a duration of 15 s, and it is bright during the first 6 s only.
The microphysical parameters describing the electron distribution are $\epsilon_e=$ 1/3, $p= 2.7$, and a varying fraction
$\zeta$ of accelerated particles.
The low magnetic energy density ($\epsilon_B$=10$^{-3}$) enhances the IC component and makes this burst an interesting
candidate for a LAT detection.
The burst has an isotropic equivalent energy $E_{\gamma,iso} = 1.35\times 10^{52}$ erg, with $1.26\times 10^{52}$ erg in the
synchrotron component and $0.09\times 10^{52}$ erg in the IC component.
The low and high-energy indices of the synchrotron spectrum are $\sim -1.1$ and $\sim -2.4$, respectively.\\

We placed the synthetic burst at a low redshift $z=0.07$ as an easy way to increase the observed flux and to produce
a large number of simulated counts in the \Fermi instruments.
As explained further below, this allowed us to characterize with high accuracy and unambiguously the properties of the
burst emission folded with the instrument responses.
As a result, the fluence of the synthetic burst is $5.4\times 10^{-4}$ \ergfluence between 10 keV and 1 MeV during the first 6 s.
This would be a very rare event among the GRBs that have been jointly detected by the GBM and the LAT, whose
fluence ranges from $5\times 10^{-8}$ \ergfluence to $\sim 3\times10^{-4}$ \ergfluence
in the same energy range \citep{LATCatalog2013}.
In order to consider more realistic situations, two other synthetic bursts were created by dividing the simulated emission
flux by 10 and 100.

In the following, the three synthetic bursts 
are denoted by $\GRBBone$, $\GRBBten$ and $\GRBBhun$ in order of decreasing flux.
We splitted the light curve of each of these bursts in three time intervals, [0 s, 1 s], [1 s, 3 s] and [3 s, 6 s].
The upper panel in the left part of Fig.~\ref{fig:multi_LC_FD} shows the 
  corresponding spectral energy distributions (SEDs) of $\GRBBten$ in addition to the SED of this burst during the total time interval [0 s, 6 s].
The lower panel shows the evolution with energy of the local photon index $\Gamma(E)$, which we calculated
numerically as the logarithmic derivative of the differential photon spectrum $F=dN/dE$ with respect to the
logarithmic energy, $\Gamma(E)=\partial \ln(F)/ \partial \ln(E)$.

\subsection{Simulation procedure}

\begin{figure*}[t!]
   \includegraphics[width=0.50\linewidth]{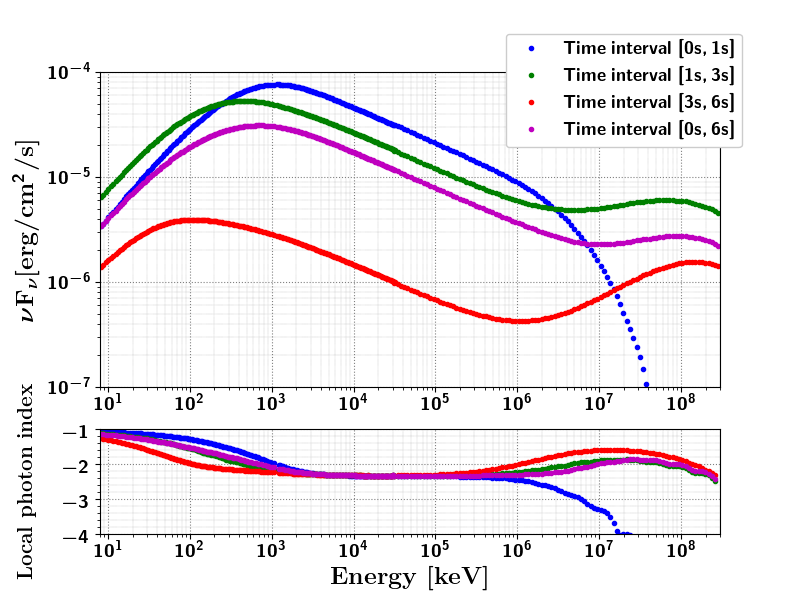}
   \includegraphics[width=0.50\linewidth]{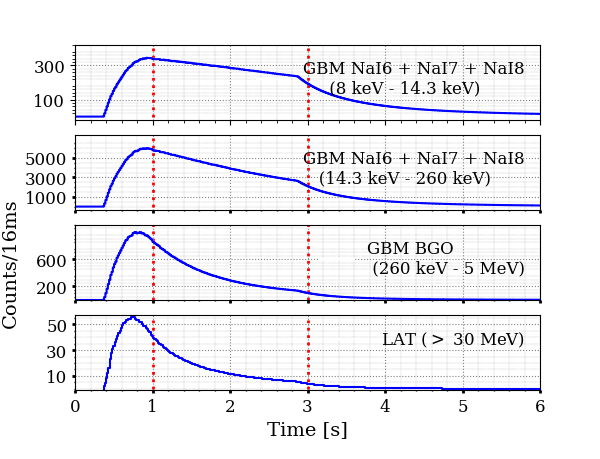}
   \caption{
   Left: Spectral energy distributions \fpc{of the synthetic burst $\GRBBten$} and local photon index in the
   four time intervals [0 s, 1 s], [1 s, 3 s], [3 s, 6 s] and [0 s, 6 s].
   Right: Multi-detector light curves of \fpc{$\GRBBten$}: summed counts in two energy ranges of GBM/NaI detectors
   (first \fpc{two} panels), in the GBM/BGO energy range (third panel) and using the largest LAT energy range (> 30 MeV) (bottom panel).
   The red dashed lines show the boundaries of the three time intervals [0 s, 1 s], [1 s, 3 s] and [3 s, 6 s].
  }
 \label{fig:multi_LC_FD}
 \end{figure*}

We simulated the signal of the synthetic bursts as it would be observed by the GBM or the LAT by performing a convolution
  of the GRB differential photon spectra $dN/dE$ with the corresponding detector response matrix (DRM).
  The DRM is defined as the detector effective area $A_\mathrm{eff}(E)$ multiplied by its energy redistribution function $D(E,E')$, where $E$ and $E'$ stand for true and measured photon energy, respectively.
The mean number of counts in the interval of measured energy [$E^\prime_\mathrm{min}$,$E^\prime_\mathrm{max}$] is given by:
\begin{equation}
N= T_\mathrm{obs} \int_{E^\prime_\mathrm{min}}^{E^\prime_\mathrm{max}} dE^\prime \int_0^{+\infty} \frac{dN}{dE}(E)\,A_\mathrm{eff}(E)\,D(E,E^\prime)\,dE
\end{equation}
where $T_\mathrm{obs}$ is the time exposure.
For this computation, we used the DRMs of the four GBM detectors (NaI6, 7, 8 and BGO1) that have seen GRB\,090926A and the
DRM of the LAT produced by the
\textit{gtrspgen}\footnote{\url{https://fermi.gsfc.nasa.gov/ssc/data/analysis/scitools/help/gtrspgen.txt}} tool available
at the \Fermi Science Support 
Center\footnote{\url{https://fermi.gsfc.nasa.gov/ssc/data/analysis/scitools/overview.html}}.
The simulation of the synthetic bursts was performed with the {\it
  XSPEC} software\footnote{\url{https://heasarc.nasa.gov/docs/xanadu/xspec}} (version 12.8.2), which generates Poisson counts of
  detected photons.
For simplicity, we did not add any background to the burst signal since it has a negligible effect owing to the large
fluence of the simulated bursts.
The multi-detector light curve of the synthetic burst $\GRBBten$ is shown in \fpc{the right part of} Fig.~\ref{fig:multi_LC_FD}.\\

%
\section{Spectral models}
\label{sec:models}
%

The \fpc{GRB spectra that we analyzed were fitted with several phenomenological functions that are commonly found
in the literature, and with a new parametric function that is built from the synthetic spectra.
All of} the functions presented below are normalized by an amplitude parameter $A$, in units of cm$^{-2}$\,s$^{-1}$\,keV$^{-1}$.

\subsection{Phenomenological models}
\subsubsection{Band function}
The \band function \citep{Band1993} is often used to fit the keV-MeV spectrum of GRBs.
\fpc{It} is composed of two smoothly-connected power laws with four parameters $A_\band$, $\alpha$, $\beta$ and
$E_p$, and it is defined as: 
\begin{equation}
 \frac{dN_\band}{dE}(E) =  A_\band \left\{
  \begin{array}{l l}
   \left(\frac{E}{100\,\mathrm{keV}}\right)^{\alpha} \exp[-\frac{E\,(2+\alpha)}{E_p}], &  E \le E_b=E_p {{\alpha-\beta}\over {2+\alpha}} \\
   \\
   \left(\frac{E}{100\,\mathrm{keV}}\right)^\beta \left[\frac{E_p}{100\,\mathrm{keV}}\frac{\alpha-\beta}{2+\alpha}\right]^{\alpha-\beta}
   \exp[\beta-\alpha] , & E > E_b=E_p {{\alpha-\beta}\over {2+\alpha}}.
\end{array}
\right.
\label{eq:band}
\end{equation}
The local photon index of this function reads:
\begin{equation}
\Gamma_\band(E) =  \left\{
\begin{array}{l l}
 \alpha - \frac{(2 + \alpha)}{E_p}E , &  E \le E_b \\
 \beta , & E > E_b.
\end{array}
\right.
\label{eq:band_slope}
\end{equation}

\subsubsection{Logarithmic parabola and variants}
The log-parabola \fpc{function} (\fpc{\logp} hereafter) has \fpc{three free parameters, i.e. one less} than the \band function.
\fpc{It} was suggested by \citep{Massaro2010ApJ} to fit \fpc{GRB spectra} and it is expressed as:
\begin{eqnarray}
\frac{dN_\logp}{dE}(E) = A_\logp \left(\frac{E}{E_0}\right)^{-\gamma-\beta \log(E/E_0)} 
\label{eq:lp}
\end{eqnarray}
\fpc{where $E_0$ is a fixed reference energy}. The local photon index is a function of the spectral parameters $\gamma$ and $\beta$:
\begin{equation}
\Gamma_\logp(E) = -\gamma-2\beta \log\left(\frac{E}{E_0}\right)
\label{eq:lp_slope}
\end{equation}

\fpc{and the \logp peak energy is $\displaystyle E_p= E_0\times 10^{\frac{2-\gamma}{2 \beta}}$.
The \logp function} is characterized by its continuous curvature, unlike the \band function.
\fpc{Its symmetric shape implies that the spectral parameter reconstruction is driven by the low-energy data, where most
  of the photon statistics is recorded.}
In order to gain some latitude at high energies, we modified the function to freeze the local photon index above a break
energy $E_b$.
\fpc{As a result, the modified logarithmic parabola, denoted by \logpm hereafter, has four free parameters:} 
\begin{equation}
\frac{dN_\logpm}{dE}(E) = A_\logpm \left\{
\begin{array}{l l}

 \left(\frac{E}{E_0}\right)^{-\gamma-\beta\,\log(E/E_0)}, &  E \le E_b \\
 \left(\frac{E_b}{E_0}\right)^{-\gamma-\beta\,\log(E_b/E_0)}\times\left(\frac{E}{E_b}\right)^{- \gamma -2\,\beta\,\log(E_b/E_0)} , & E > E_b
\end{array}
\right.
\label{eq:lp1} 
\end{equation} 

We introduced a similar modification at low energies,
\fpc{which relaxes the dependency of the spectral fit around the peak energy on the low-energy data. The corresponding modified
  logarithmic parabola, denoted by \logpmm hereafter, has five free parameters:}
\begin{equation}
\frac{dN_\logpmm}{dE}(E) = A_\logpmm \left\{
\begin{array}{l l}

\left(\frac{E_b^\prime}{E_0}\right)^{-\gamma-\beta\,\log(E_b^\prime/E_0)}\times\left(\frac{E}{E_b^\prime}\right)^{- \gamma -2\,\beta\,\log(E_b^\prime/E_0)}  , &  E \le E_b^\prime \\
\left(\frac{E}{E_0}\right)^{-\gamma-\beta\,\log(E/E_0)}  , & E_b^\prime \le E \le E_b \\
\left(\frac{E_b}{E_0}\right)^{-\gamma-\beta\,\log(E_b/E_0)}\times\left(\frac{E}{E_b}\right)^{- \gamma -2\,\beta\,\log(E_b/E_0)}  , & E > E_b
\end{array}
\right.
\label{eq:lp2}
\end{equation} 

\subsubsection{(Broken) power law with exponential cutoff}

For the spectral analysis of GRB\,090926A presented in Sect.~\ref{sec:ana_090926A}, which extends to the LAT energy range, we adopted either a power law with exponential cutoff (\cutpl) or a
broken power law with exponential cutoff (\cutbpl). \fpc{The \cutpl function is expressed as:}
\begin{equation}
  \frac{dN_\cutpl}{dE}(E) = A_\cutpl\left(\frac{E}{E_0}\right)^{\lambda} \exp\left(-\frac{E}{E_{f}}\right)
  \label{eq:cutpl}
\end{equation}

\fpc{which has three free parameters $A_\cutpl$, $\lambda$ and the folding energy $E_{f}$ of the exponential cutoff, and a fixed
reference energy $E_0$.}
\fpc{The \cutbpl function is expressed as:}  
  \begin{equation}
\frac{dN_\cutbpl}{dE}(E) = A_\cutbpl \left\{
\begin{array}{l l}
   \left(\frac{E}{E_0}\right)^{\gamma_0}\exp\left(-\frac{E}{E_f}\right), &  E \le E_b \\
\\
    \left(\frac{E_b}{E_0}\right)^{\gamma_0}\left(\frac{E}{E_b}\right)^\gamma \exp\left(-\frac{E}{E_f}\right), & E > E_b
\end{array}
\right.
\label{eq:cutbpl}
\end{equation}
where $E_b$ is the break energy, $\gamma_0$ and $\gamma$ are the photon index below and above $E_b$, \fpc{respectively}.
  \fpc{As explained in Y17, the} break \fpc{energy} and the photon spectral index below the break have been
  fixed to $E_b = 200$ keV and $\gamma_0= +4$ in order to \fpc{cancel the contribution of the power-law component at low
    energies, as for instance expected from an inverse Compton spectral component that would extend the synchrotron
    spectrum at high energies only.
    As a result, the \cutbpl function has the same number of free parameters as the \cutpl function.} 

\subsection{The \issm spectral model}

In order to build a function that is representative of the \fpc{synchrotron spectral component of the synthetic bursts}, we fitted their local photon index as a function of energy with the following parameterization:
\begin{equation}
\Gamma(E) = \frac{\partial \ln(F)}{\partial \ln(E)} = - a + \frac{b}{ E + c }
\label{eq:slopes_func}
\end{equation}
where $a$, $b$, $c$ are \fpc{free parameters}.
\fpc{This parameterization} adequately fits the local photon index of $\GRBBone$ in the four time intervals as shown in
the left panel of Fig.~\ref{fig:slopes_function_fit}.
\magenta{The right panel of this figure shows that it is also suitable for different configurations of the
  model presented in BD14.
Note that the synthetic bursts using various assumptions for the microphysics in the emission region do not have
    the same low-energy photon index: $\sim -1.5$ for case A as expected for the standard fast cooling
    synchrotron spectrum, and $-1.1$ to $-0.75$ for case B.} 
Integrating Eq.~\ref{eq:slopes_func}, one gets:
\begin{figure}
\centering
\hbox{
  \includegraphics[width=0.47\linewidth]{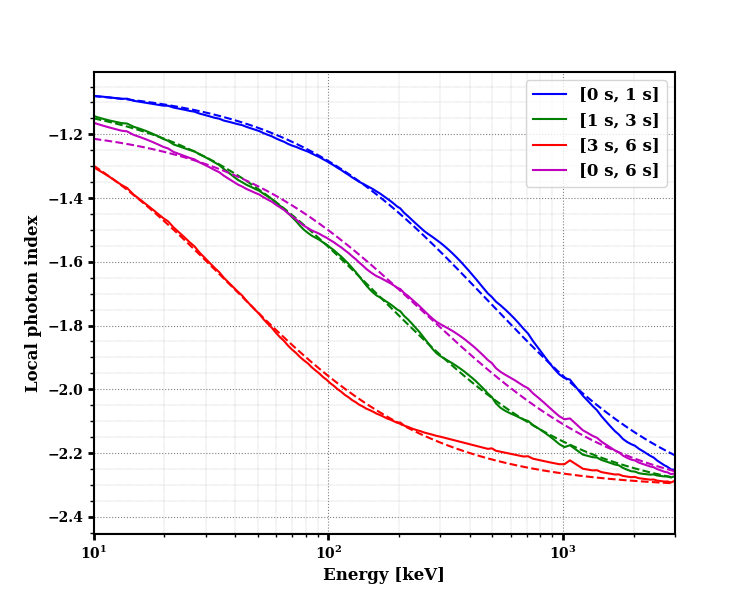}
  \includegraphics[width=0.47\linewidth]{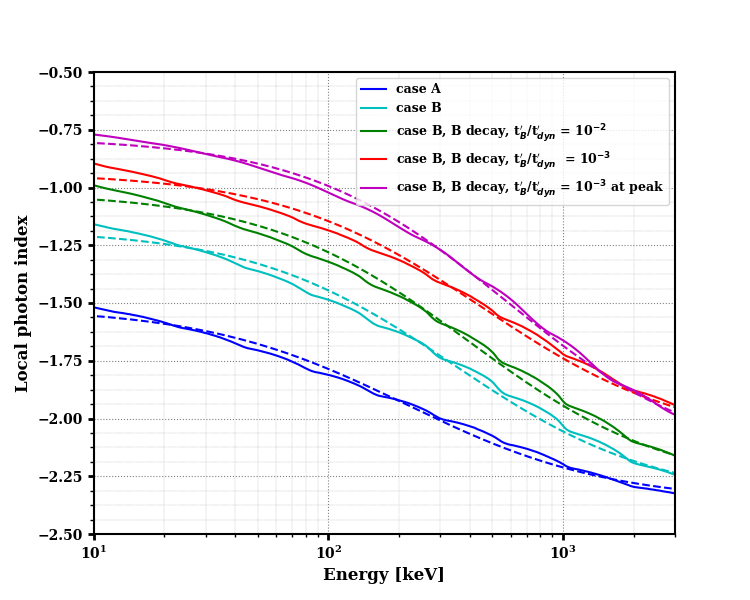}
}
\caption{Left: Least-square fit \fpc{(dashed lines)} of Eq.~\ref{eq:slopes_func} to the local photon index \fpc{(thick lines)} in the \fpc{keV-MeV}
  range for the four time intervals of the synthetic burst $\GRBBone$.
  \magenta{Right: Least-square fit (dashed line) to the local photon index (solid line) in the keV-MeV range for several synthetic
    bursts: time integrated spectrum of two reference cases presented in BD14, case A (blue) and B (cyan) with a varying
    fraction of accelerated electrons and $p=2.7$. Case A corresponds to the most standard fast cooling synchrotron
    spectrum, and case B to a modified synchrotron spectrum affected by inverse Compton scaterring in the Klein-Nishina
    regime. 
    In addition, preliminary calculation of the same case B taking into account a magnetic field decay in the emission
    region are also presented (taken from Daigne \& Bo{\v s}njak, in preparation) with a time scale of the decay $t’_\mathrm{B}/t’_\mathrm{dyn}=10^{-2}$ (green) or $10^{-3}$
    (red), where $t’_\mathrm{dyn}$ is the dynamical timescale. Finally, the same case is also shown for a time interval of
    $0.25\,\mathrm{s}$ around the peak of the light curve (magenta).}
}
\label{fig:slopes_function_fit}
\end{figure}

\begin{equation}
  F(E) = F(E_r) \exp \left[-a \ln\left(\frac{E}{E_r}\right)\right] \exp\left[\frac{b}{c}\ln\left(\frac{E\,(E_r+c)}{E_r\,(E+c)}\right)\right]
\label{eq:integrated_func}
\end{equation}
\fpc{where the reference energy $E_r$ is related to the constant of integration. 
From this parameterization, the asymptotic spectral indices towards low and high energies can be easily obtained as $\alpha$=
$\frac{b}{c} - a$ and $\beta = -a$, respectively.
Finally, defining the SED peak energy $E_p$ as the solution of $\Gamma(E)=-2$:}
\begin{equation}
  E_p=-c\left(\frac{2+\alpha}{2+\beta}\right),
\label{eq:issm_ep}
\end{equation}
\red{one can rewrite Eq.~\ref{eq:integrated_func} to obtain a new expression, denoted by \issm
hereafter:
\begin{equation}
\frac{dN_\issm}{dE}(E) = \frac{A_\issm}{\left[1 - \frac{E_p}{E_r} \left(\frac{2+\beta}{2+\alpha}\right)\right]^{\beta-\alpha}}\times \left(\frac{E}{E_r}\right)^\alpha \left[\frac{E}{E_r} - \frac{E_p}{E_r} \left(\frac{2+\beta}{2+\alpha}\right)\right]^{\beta-\alpha}
\label{eq:issm}
\end{equation}
which has four parameters $A_\issm$, $\alpha$, $\beta$, $E_p$. It is important to note that $E_r$ is a fixed reference
energy which is chosen as the energy at which the flux normalization is defined:
\begin{equation}
  \frac{dN_\issm}{dE}(E_r)=A_\issm.
  \label{eq:issm_er}
\end{equation}
In other words, different choices of $E_r$ only affect the flux normalization parameter $A_\issm$ and not the shape of the
\issm function.
The local photon index is given by:
\begin{equation}
\Gamma_\mathrm{\issm}(E) = \alpha + (\beta - \alpha)\frac{E}{E- E_p\left(\frac{2+\beta}{2+\alpha}\right)}.
\label{eq:issm_slope}
\end{equation}
The four parameters of the \issm (flux normalization, SED peak energy and asymptotic slopes) resemble those of the \band function. The local photon index $\Gamma_\mathrm{\issm}(E)$ decreases continuously with energy and the \issm function is continuously curved unlike the \band function, and unlike simplified versions of the synchrotron model based on pure power-law energy distributions of the accelerated electrons.
In the framework of our internal shock synchrotron model, the spectral curvature arises essentially from the superposition of instantaneous
electron synchrotron spectra which vary significantly within the time intervals considered by the observer, owing to the dynamical
evolution in the shock region. While we only tested the \issm function on a simple, single-pulse
    burst, we are confident that it can also represent complex burst spectra resulting from various distributions of the
    Lorentz factor. Indeed, in most cases, complex bursts can be interpreted in terms of a succession of individual pulses
    so that time-dependent spectra of complex bursts can likely be fitted in the same way. Moreover, BD14 actually
    explored in detail how the observed emission of a single pulse depends on the various physical parameters of the
    internal shock model. Their study shows that the assumptions about the dynamics (Lorentz factor,
    kinetic energy flux, etc.) affect the pulse light curve but have little effect on the shape of the spectrum.
}

%
\section{Spectral analysis of the synthetic bursts}
\label{sec:ana_sim}
%
We first focused our study of the three synthetic bursts in the GBM energy range (8 keV to 40 MeV).
The four phenomenological functions and the \issm function were used to fit the spectra of the synthetic
bursts in the four time intervals [0 s, 1 s], [1 s, 3 s], [3 s, 6 s] and [0 s, 6 s]
using the \textit{XSPEC} software.
The reference energy $E_0$ in Eqs.~\ref{eq:lp}, \ref{eq:lp1} and \ref{eq:lp2} was fixed to $500$ keV.
For simplicity, the reference energy in Eq.~\ref{eq:issm}, which relates to the flux normalization,
was fixed to the true peak energy of the synthetic spectra: $E_r$ =
$1150$, $478$, $114$ and $745$ keV for the time intervals [0 s, 1 s], [1 s, 3 s], [3 s, 6 s] and [0 s, 6 s],
respectively.
To compare the quality of the fits between the different functions, we defined the following quality factor $Q$ that mimicks a reduced $\chi^2$:

\begin{equation}
Q= \frac{1}{n-n_\mathrm{par}}\sum \limits_{i=1}^n \left(\frac{\Gamma(E_i) - s(E_i)}{\sigma_i}\right)^2
\label{eq:quotient}
\end{equation}
where $\Gamma(E_i)$ is the local photon index of the fitted function and $s(E_i)$ is the true index of the
synthetic spectrum at energy $E_i$.
The error $\sigma_i$ on $\Gamma(E_i)$ is obtained by propagating the errors of the $n_\mathrm{par}$ fitted function parameters.
\\

\begin{table*}[!t]
  \centering
  \caption{
    \cstat values of the spectral fits of the three synthetic bursts,
    performed with the five functions: \band, \logp, \logpm, \logpmm and \issm.
  }%
  \begin{tabular}{llccccc}
 \hline
 \hline
Synthetic GRB &  Model & DOF   &   \multicolumn{4}{c}{\cstat for time intervals:}\\
 &   &   &   [0-1]\,s&  [1-3]\,s& [3-6]\,s &[0-6]\,s\\
\hline
\multirow{6}*{$\GRBBone$}  &   \band &473   &       603   &       906   &       486   &       1403\\
 & \logp &474   &       768   &       677   &       615   &       631\\
 & \logpm &473  &       768   &       569  &       615   &       589\\
 & \logpmm &472   &       526   &       540   &       558   &       570\\
 &   \issm &473  &        486   &       498   &       452  &       638\\
 \hline
\multirow{6}*{$\GRBBten$} &  \band &473  &       458   &       539  &       447   &       578\\
 &  \logp &474   &       470   &       559   &       455   &       469\\
 &  \logpm &473  &       470   &       534   &       455   &       469 \\
 &  \logpmm &472   &       439   &       525   &       455   &        466  \\
 &   \issm &473  &        441   &        523  &        443   &       484 \\
 \hline
\multirow{6}*{$\GRBBhun$}&   \band &473   &       465   &       446   &       359   &       461  \\
 & \logp &474   &       464   &       445   &       364   &       447  \\
 & \logpm &473  &       464   &       445   &       364   &       447 \\
 & \logpmm &472   &       462   &       442   &       376   &       447  \\
 &  \issm &473  &        463   &       444  &       360   &       449  \\
 \hline
  \end{tabular}
  \label{table:cstat_3classes}
\end{table*}

\begin{figure*}[!t]
  \centering
  \hbox{
  \includegraphics[width=0.5\linewidth]{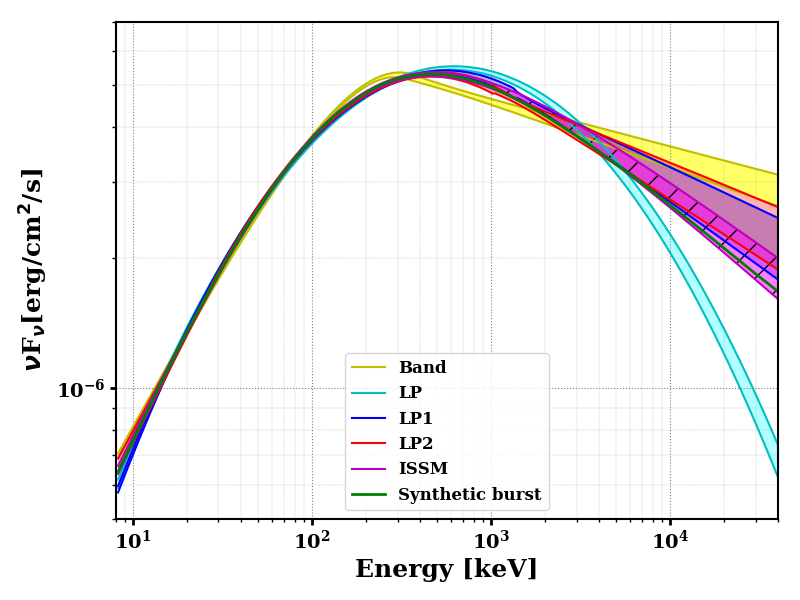}
  \includegraphics[width=0.5\linewidth]{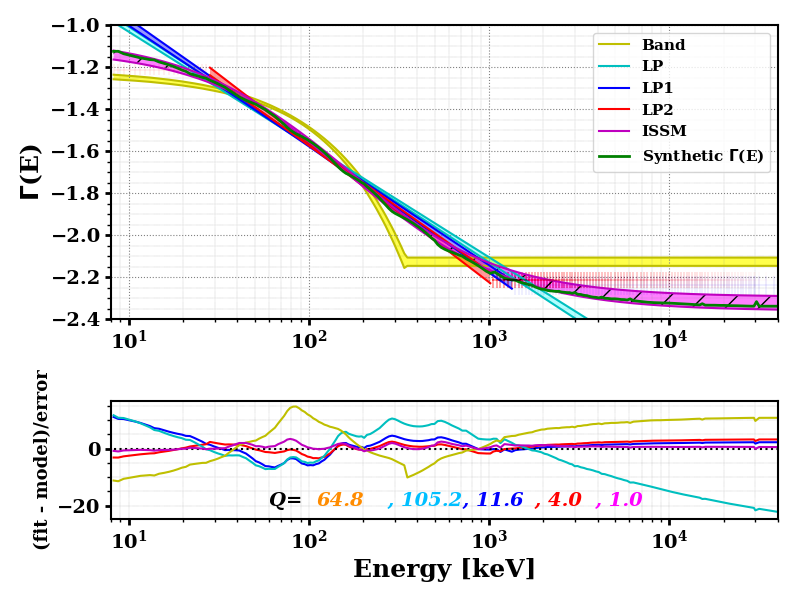}
  }
  \caption{
  Left: SEDs of the $\GRBBten$ synthetic burst in the time interval [1 s, 3 s], from fits with the five spectral functions. The fit with the \issm function is presented by the hatched magenta line.
  Right: Local photon index as a function of the photon energy.
  The fit quality factor $Q$ of the five functions is given in the bottom panel.
  }
  \label{fig:nufnu_Index_1_3_flux_10_allModels}
\end{figure*}

The spectral analyses were performed using the Castor fit statistic\footnote{See
    \url{https://heasarc.nasa.gov/docs/xanadu/xspec/xspec11/manual/node57.html}.} (\cstat) for Poisson
distributed total counts of the burst.
The \cstat values obtained from the fits of the three synthetic burst spectra are reported in Table~\ref{table:cstat_3classes}. 
The \issm function has the lowest \cstat value in most of the time
intervals especially for the synthetic burst with the highest flux value i.e. $\GRBBone$. For $\GRBBhun$, all functions yield similar \cstat values, meaning  
  that the fits are of similar quality owing to the low photon statistics for this faint burst.
Fig.~\ref{fig:nufnu_Index_1_3_flux_10_allModels} shows the SEDs and local photon index of the $\GRBBten$ burst in the
time interval [1 s, 3 s], as obtained from the fits with the five spectral functions. As can be seen from this figure, both 
the SED and the local photon index 
are not reproduced by the \band function fit, 
in particular around and above the peak energy.
The fit quality of the \logp function is even worse owing to the linear dependency of its local photon index with energy, which is
not adequate at low and high energies.
The \logpm and \logpmm functions provide better fits and their parameters are not
constrained for the three bursts in all the time intervals.
Finally, Fig.~\ref{fig:nufnu_Index_1_3_flux_10_allModels} shows that the \issm function has the lowest $Q$ value
among all fitted functions, which is 
expected from this model that has been built directly from the synthetic spectra.\\



By nature, the \issm function reproduces the keV-MeV spectra of the synthetic bursts simulated with the internal shock
synchrotron model.
It has the same number of free parameters as the \band function, which is commonly used to fit the prompt high-energy
spectrum of GRBs.
Therefore, before applying these functions to real GRB observations (see Sect.~\ref{sec:ana_gbm}), it is worth comparing
their shapes in detail.
Tables \ref{table:Band_fit_flux}, \ref{table:Band_fit_flux10}, and \ref{table:Band_fit_flux100} in appendix show the parameters of the
\band and \issm fits to $\GRBBone$, $\GRBBten$ and $\GRBBhun$, respectively.
The asymptotic low-energy index $\alpha$ of the \issm function is found to be larger than that of the \band function,
while the high-energy index $\beta$ is smaller.
Interestingly, the peak energies of the synthetic bursts are estimated with much greater accuracy with the \issm
  function than with the \band function, which underestimates them by $\sim$36$\%$.
Furthermore, we compared the spectral width of the two functions, following \cite{Yu2015A&A} who proposed a method to
calculate the SED sharpness around its peak energy. We did not consider the alternate measure of
    the spectral width proposed by \cite{Axelsson2015}, which is defined as $W=\log(E_2/E_1)$ where $E_1$ and $E_2$ are the energy bounds of the SED full width at half maximum. The spectral sharpness angle defined by \cite{Yu2015A&A} is computed from the triangle defined by the vertices at $E_p/10$, $E_p$ and $3E_p$.
To compute this angle and its asymmetrical errors accurately, we performed Monte-Carlo simulations using the fit
parameters and their covariance matrix, assuming that their distribution is a multivariate Gaussian.
This process was repeated $1000$ times for each time interval and for each of the two bright synthetic bursts
$\GRBBone$ and $\GRBBten$.
The spectral sharpness angle was chosen as the maximum probability value (MPV) of the distribution obtained from the $1000$
realizations.
The errors on the angle were calculated from the $68$\% confidence intervals on each side of the MPV.
The results of this analysis are reported in Table~\ref{table:angles_2bursts}, which confirms that the \issm
function reproduces the spectral width of the synthetic bursts better than the \band function.\\


\begin{table*}[t!]
  \centering
  \caption{
  Spectral sharpness angle (in degrees) from the \band and \issm fits to the synthetic bursts $\GRBBone$ and $\GRBBten$.
  }
  \begin{tabular}{llcccc}
\hline
\hline
Synthetic GRB &  Model  &   \multicolumn{3}{c}{Time interval}\\
 &     &   [0-1]\,s&  [1-3]\,s& [3-6]\,s &[0-6]\,s\\
\hline
\multirow{5}*{  $\GRBBone$}  &   Synthetic    &   142.3              &   145.9              &  145.5  &   148.9 \\
                                        &  \band              &   137.7$_{-0.2}^{+0.4}$     &   140.0$_{-0.1}^{+0.2}$    &  142.7$_{-0.3}^{+0.6}$    & 142.0$_{-0.1}^{+0.2}$     \\
                                        &    \issm     &   143.5$_{-0.5}^{+0.5}$     &   145.7$_{-0.3}^{+0.3}$    &  145.8$_{-1.0}^{+1.3}$   &  148.6$_{-0.3}^{+0.4}$    \\
 \hline
\multirow{5}*{ $\GRBBten$} &  Synthetic &  142.3    &   145.9   &  145.5  &   148.9 \\
                                             &   \band        &  139.1$_{-1.2}^{+0.9}$      &  139.9$_{-0.4}^{+0.7}$      & 145.1$_{-1.4}^{+1.8}$     &  142.6$_{-0.6}^{+0.4}$    \\
                                             &   \issm &  143.6$_{-1.3}^{+1.8}$      &  146.5$_{-1.2}^{+1.1}$      &  150.6$_{-4.7}^{+4.7}$   &   148.9$_{-0.8}^{+1.3}$   \\
 \hline
  \end{tabular}
  \label{table:angles_2bursts}
\end{table*}

\begin{figure*}[!t]
  \centering
  \caption{
  Left: Fit of the count spectrum of the synthetic burst $\GRBBone$ with the \issm + \cutpl model in the time interval [1 s - 3 s]. 
  Right: Fit of the \issm and \band models to the GRB\,150403913 spectrum oberved by the GBM.
  }
  \hbox{
  \includegraphics[width=0.5\linewidth]{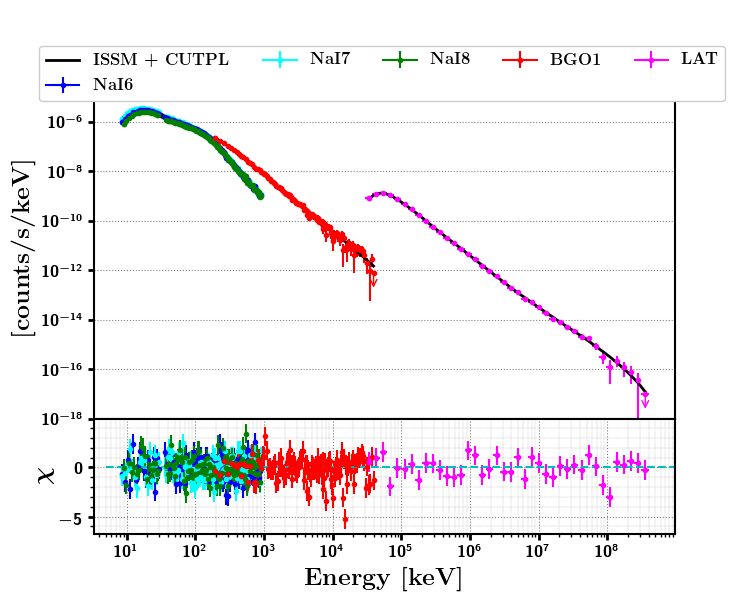}
  \includegraphics[width=0.5\linewidth]{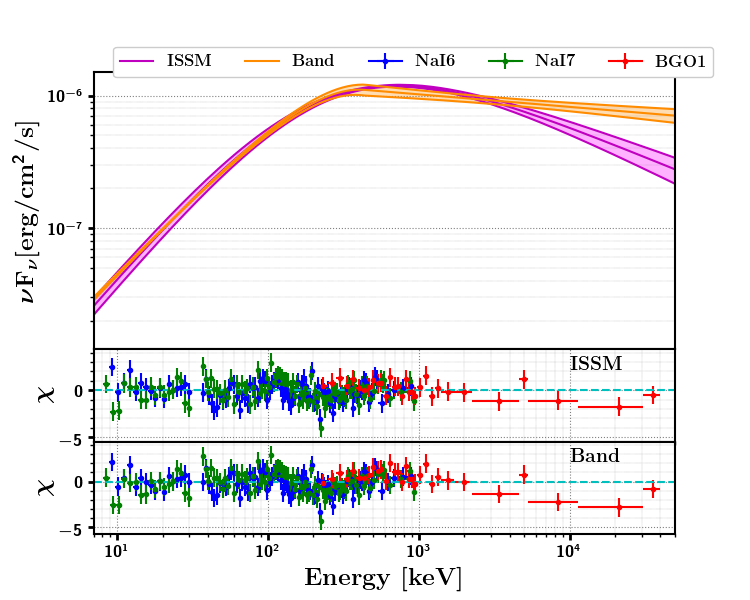}
  }
  \label{fig:keV_GeV_1_3_3_6_CUTPL_CUTBPL}
\end{figure*}

For the sake of completeness, we carried out broadband spectral analyses of the brightest synthetic burst
($\GRBBone$) in the two time intervals [1 s - 3 s] and [3 s - 6 s], where the inverse Compton spectral component is
  prominent.
  We used the \cutpl model to fit this high-energy spectral component and fixed the reference energy $E_0$ to $10$ GeV in
  Eq.~\ref{eq:cutpl}. This value is close to the decorrelation energy and thus minimizes the correlation between the \cutpl parameters.
Despite its brightness in the LAT energy range, the inverse Compton component of $\GRBBone$ peaks at $\sim 100$ GeV, where
few simulated events are recorded. We multiplied artificially the LAT effective detection area by $100$ to get rid of
these statistical limitations and to check whether the adopted model is able to capture all features in the internal shock model spectra.
The fit results obtained with the \issm + \cutpl model in the two time intervals are reported in Table
\ref{table:IAP_fit_flux_keV_GeV_cutpl_CUTBPL} and shown in the left panel of Fig.~\ref{fig:keV_GeV_1_3_3_6_CUTPL_CUTBPL} for the time interval [1 s - 3 s].
The fit residuals and reduced \cstat values clearly show the excellent quality of the fits and the ability of the
  \issm + \cutpl model to reproduce the broadband shape of the synthetic spectra.

\begin{table}[t!]
  \centering
  \caption{
  Results of the \issm+\cutpl fits to the synthetic burst $\GRBBone$ during the time intervals [1 s - 3 s] and [3 s - 6 s].
  }
  \begin{tabular}{llccc}
\hline  
\hline  
Spectral          &  Fit results  &   \multicolumn{2}{c}{Time interval}\\
component         &               &     [1-3]\,s     & [3-6]\,s  \\
\hline
 \issm& $E_p$ (keV) &     458 $\pm$ 4    & 119  $\pm$     3\\
&\bf $\alpha$ &  -1.09 $\pm$ 0.01 &  -1.03   $\pm$    0.05 \\
&\bf $\beta$   &  -2.354 $\pm$ 0.003  & -2.321   $\pm$    0.004 \\
&A$_{MeV}$ (\unit) & 0.145 $\pm$ 0.001  &  0.193   $\pm$    0.001 \\
 \hline
\cutpl &  $\lambda$ & -1.51 $\pm$  0.06  & -1.28       $\pm$ 0.07 \\
&E$_f$ (GeV) &  165 $\pm$ 71 &  172  $\pm$  91 \\
&A$_{GeV}$ (\unit) & (167.4 $\pm$  11.5)$\times$10$^{-13}$  & (39.0    $\pm$   3.5)$\times$10$^{-13}$\\
 \hline
\issm + \cutpl& \cstat/DOF & 533/510  &   502/510  \\
\hline
  \end{tabular}
  \label{table:IAP_fit_flux_keV_GeV_cutpl_CUTBPL}
\end{table} 

%
\section{Application to GBM bursts}
\label{sec:ana_gbm}
%

\subsection{GRB sample and data selection}

According to the results presented in Sect.~\ref{sec:ana_sim}, a large number of counts is required to
distinguish the different spectral model based on their fit quality.
For this reason, we selected a sample of bursts detected by the GBM with an energy fluence larger than $10^{-5}$
erg\,cm$^{-2}$ (from 10 to 1000 keV), namely comparable to those of the $\GRBBone$ and $\GRBBten$ synthetic bursts.
Like in Sect.~\ref{sec:ana_sim}, we first focused our study on the sub-MeV spectral component, discarding the bursts
that have additional components at low or high energies.
This includes the GRBs with a low-energy excess that has been interpreted as a possible thermal component (GRB\,090424, GRB\,090820
\citep{Tierney2013AA}, GRB\,090902B \citep{090902B2009}, GRB\,090926A \citep{Guiriec2015ApJ...807..148G}, GRB\,100724B
\citep{Guiriec2011}, GRB\,110721 \citep{Axelsson2012}), the GRBs with an extra high-energy power-law component
(GRB\,080916C \citep{LATCatalog2013}, GRB\,090902B \citep{090902B2009}, GRB\,090926A
\citep{Ackermann2011ApJ...729..114A}) or with a strong spectral evolution (GRB\,081215A \citep{Tierney2013AA}).
The bursts whose spectra are best fitted by a simple power law in the GBM spectral catalog\footnote{\url{https://heasarc.gsfc.nasa.gov/W3Browse/fermi/fermigbrst.html}} \citep{Gruber2014} were
also excluded.
Beside these 15 GRBs, we eliminated the bursts which have been seen by NaI detectors
with a separation angle between the detector axis and the source larger than $60^{\circ}$.
As a result, we selected $74$ GBM GRBs in the first eight observing years, which are listed in
Table~\ref{tab:catalog:Band_IAP} in appendix.
More than half of them ($41$) are best fitted by the \band function in the GBM spectral catalog \citep{Gruber2014}.
Another fair fraction of bursts from this catalog ($24$) are best fitted by a power law with an exponential cutoff. This model is a special case of the \band function that is obtained for a very steep high-energy index (i.e., $\beta$ tends to $-\infty$
  and $E_b$ to $+\infty$ in Eq.~\ref{eq:band}).
The remaining $9$ GRBs were found to be best fitted by a smoothly broken power law by \cite{Gruber2014}, which is
characterized by a flexible SED width around its peak energy. The data are loaded from the FSSC GBM data \footnote{\url{https://fermi.gsfc.nasa.gov/ssc/data/access/gbm/}} using gtburst tool \footnote{\url{https://fermi.gsfc.nasa.gov/ssc/data/analysis/scitools/gtburst.html}}. The spectral analyses where performed during the T90 defined in GBM catalog \cite{Gruber2014}.
For each GRB of the sample, we selected one BGO detector with a separation angle less than $90^{\circ}$ and a maximum of
three NaI detectors that have seen the burst with a separation angle less than $60^{\circ}$.

\subsection{Model comparison}

\begin{figure*}[!t]
  \centering
  \hbox{
    \includegraphics[width=.6\linewidth]{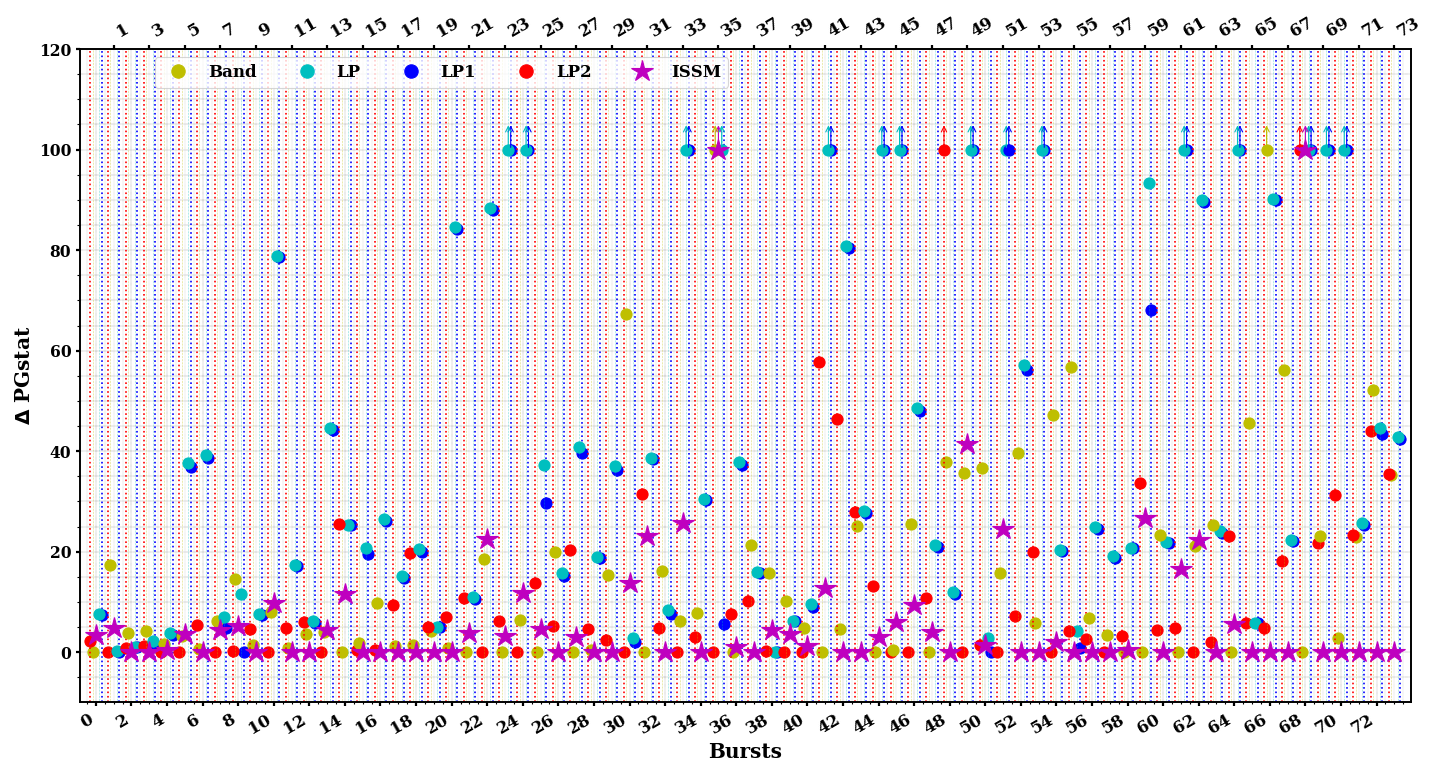}
    \includegraphics[width=.4\linewidth]{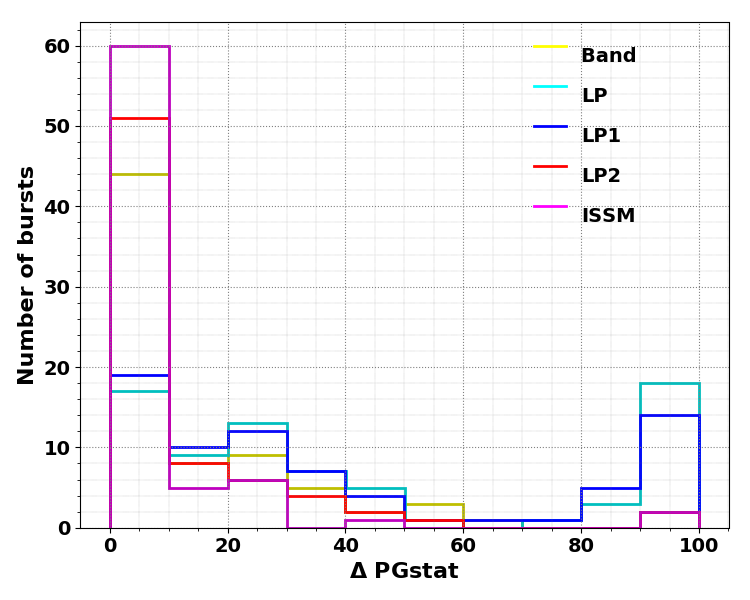}
  }
  \caption{
    Left: Difference in \pgstat of the five models with respect to the model with the lowest \pgstat, for every GRB
      displayed with increasing SNR (increasing from the left to the right).
    For each GRB, the five model markers are displayed within two vertical (red and blue) lines.
      By definition, the model with the lowest \pgstat is always placed on the zero horizontal line.
    Lower limits at the top of the figure stand for models with a $\Delta$\pgstat larger than $100$.
    Right: Distribution of $\Delta$\pgstat for the five models.
    }
  \label{fig:deltaPGstat_distribution_real_data}
\end{figure*}

\fpc{We performed a spectral analysis} of the \fpc{$74$} selected GRBs with the {\it XSPEC} software and for the five spectral models: \band, \logp, \logpm, \logpmm and
\issm. The reference energies $E_0$ in Eqs.~\ref{eq:lp}, \ref{eq:lp1}, \ref{eq:lp2} and $E_r$ in
Eq.~\ref{eq:issm} of the \logp, \logpm, \logpmm and \issm functions, \fpc{were} fixed to $500$ keV.
We used the ``Poisson-Gauss'' fit statistic\footnote{See \url{https://heasarc.gsfc.nasa.gov/xanadu/xspec/manual/XSappendixStatistics.html}} (\pgstat hereafter), which is suitable for GRB spectral analysis, where
  the observed data counts are Poisson distributed in the energy channels, while background counts have been estimated
  beforehand from pre- and post-burst data and are assumed to follow a Gaussian distribution.
The case of GRB\,150403913 is shown in the right panel of Fig.~\ref{fig:keV_GeV_1_3_3_6_CUTPL_CUTBPL} for the
  \issm and \band fits.
The left panel of Fig.~\ref{fig:deltaPGstat_distribution_real_data} shows the increase of \pgstat of the five models with respect to the model which has the lowest \pgstat (``reference model'' hereafter).
In this panel, the GRBs are displayed in order of increasing signal-to-noise ratio (SNR), which is defined for each GRB as:
  \begin{equation}
\mathrm{SNR}=\sum \limits_{i=1}^N(c_i-b_i)\,/\sqrt{\sum \limits_{i=1}^N b_{i}}
\label{eq:snr}
  \end{equation}
where $c_{i}$ (resp. $b_{i}$) are the total (resp. background) counts recorded by the $N$ NaI detectors that have detected the burst.
The right panel of Fig.~\ref{fig:deltaPGstat_distribution_real_data} shows the resulting distribution of
  $\Delta$\pgstat for the five models.
The \issm function has the lowest \pgstat, namely it is the reference model, for half of the 
GRBs in Fig.~\ref{fig:deltaPGstat_distribution_real_data}. Since the ISSM function shows the lowest value of \pgstat in half of the cases, it is taken as a reference (level 0 on the bottom of Fig.~\ref{fig:deltaPGstat_distribution_real_data}) and the other models are displayed accordingly.
The first GRBs with the minimum SNR values in this figure have comparable \pgstat values for the five spectral models and the $\Delta$\pgstat increases
with SNR as expected, since the models can be more easily distinguished from each other with a larger event statistics.\\

\begin{figure*}[!t]
  \centering
  \hbox{
  \includegraphics[width=0.5\linewidth]{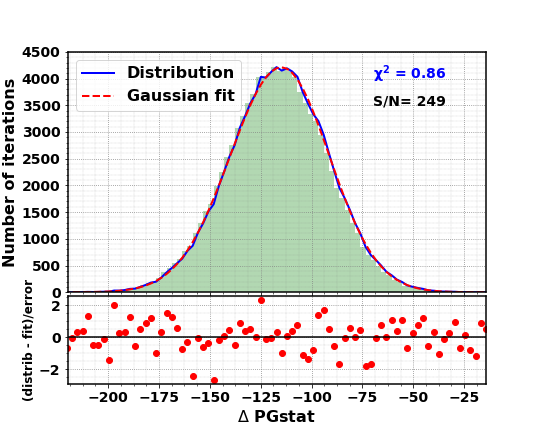}
  \includegraphics[width=0.5\linewidth]{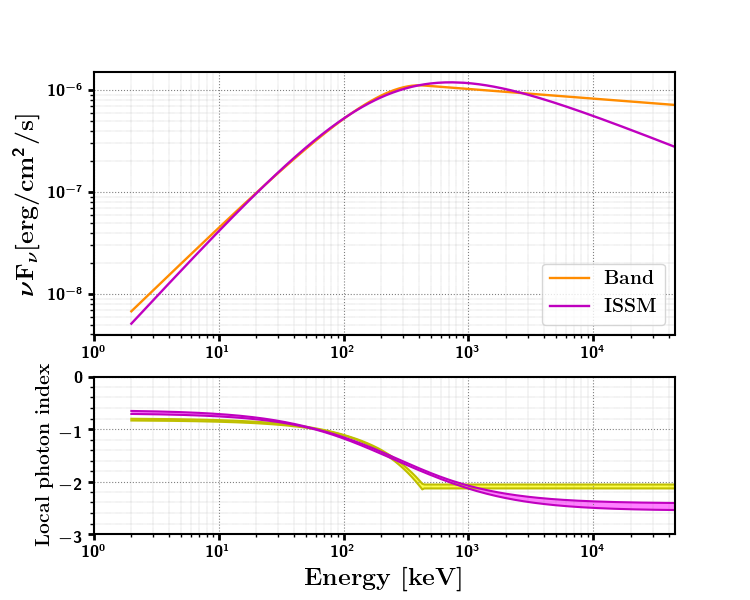}
  }
  \caption{
    Left: Fit of an asymmetric Gaussian function to the distribution of $\Delta$\pgstat between the \band and the
    \issm function for the medium SNR case. The bottom panel shows the ratio of the difference between the
    histogram and its fit over the error ($\sqrt N$ in each bin).
    Right: 
    \fpc{Spectral energy distribution and} local photon index of a representative \fpc{burst (GRB\,150403913)} with the
    \band and \issm functions.
  }
  \label{fig:asymetric_gaussian_fit}
\end{figure*}

To compare the fitted models with each other, we used the $\Delta$\pgstat as a likelihood ratio test \citep{NeymanPearson}.
In case of nested models, where the model parameterization in the null hypothesis is a special case of that
in the alternative hypothesis, the $\Delta$\pgstat is expected to follow a $\chi^{2}$ distribution with $k$ degrees
of freedom in the large sample limit, where $k$ is the number of additional parameters between the two
models~\citep{wilks1938}.
Since several of the models that we considered are not nested, and because the large sample limit is not reached in all
energy channels of the GRBs in our sample, one should compute the $\Delta$\pgstat probability density function
for each GRB and each pair of models by simulating a large number of spectra.
Given the \fpc{vast} number of cases, we focused on the \band and \issm functions, in the two cases of a low or a medium value of the SNR.
We performed Monte-Carlo simulations for two cases \fpc{in our sample}, GRB\,100910A (SNR=141) and GRB\,110921A (SNR=249),
considering the \band function as the null hypothesis.
\fpc{We used the {\it XSPEC} software to simulate} 10$^{5}$ \band spectra for the duration of each GRB, using the DRM
  \fpc{and background files} of the GBM detectors that have seen the burst with the best viewing angle.
  All the simulated spectra were then fitted with the \band and \issm functions.
  The resulting distribution \fpc{of} $\Delta PG_\mathrm{stat}=PG_{\mathrm{stat},\band}-PG_{\mathrm{stat},\issm}$ \fpc{for} GRB\,110921A
  is shown in the left panel of Fig.~\ref{fig:asymetric_gaussian_fit}.
The fit of this distribution with an asymmetric Gaussian function and its extrapolation allowed us to compute the
  $\Delta$\pgstat limit beyond which the probability that a statistical fluctuation yields a better fit with the \issm
  function than with the \band function is smaller than $10^{-6}$ (approximately $5$ Gaussian standard deviations). 
The limit was found to be $\Delta$\pgstat$=20$ for a low SNR and $3$ for a medium SNR, beyond which the null hypothesis
(i.e., the \band function) must be rejected.
Because it was complicated and time consuming to determine a limit for each GRB and each pair of models, we adopted a
\fpc{common} limit of $\Delta$\pgstat$=10$ in all situations.


\fpc{As a result,} this study revealed that the \issm model is the reference model for $36$ GRBs, $19$ of which are
equivalently fitted by the \band function.
On the contrary, the \band function has the lowest \pgstat value for $16$ GRBs, $10$ of which are \fpc{equivalently} fitted by the
\issm function.
Concerning the other three models, only the \logpmm \fpc{showed} good performance.
It is the reference model for $18$ GRBs, and globally as good as the \band model, though with one more parameter.
All in all, the \issm function is a good spectral model for $81$\% ($60/74$) of the GRBs in our sample, namely in these cases it is the reference
  model or it is close enough to it in terms of \pgstat.
  The \band function was found to be a good spectral model for a smaller fraction ($59$\%) of the GRB
  sample ($44/74$), similar to the \logpmm function ($65$\%), versus only $\sim$$20$\% for the \logpm and \logp functions. 
  It must be noted that these performances would improve for more common and less fluent bursts with lower signal-to-noise ratios.

\subsection{\band and \issm spectral parameters}
\begin{figure*}[!t]
  \centering
\hbox{
  \includegraphics[width=0.5\linewidth,height=6.4cm]{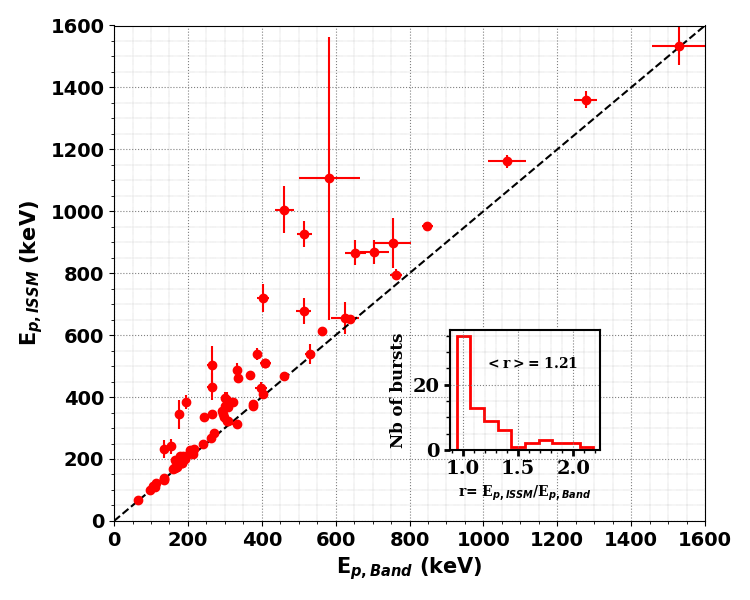}
  \includegraphics[width=0.5\linewidth,height=7cm]{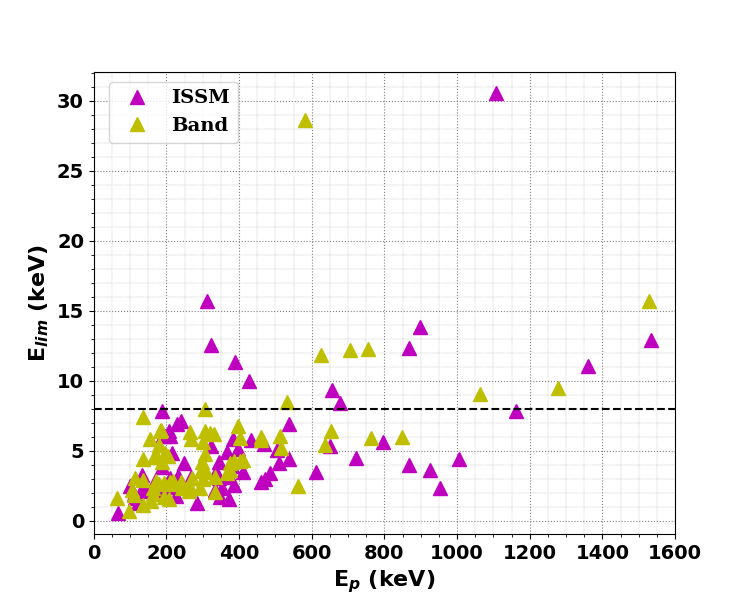}
  }
  \caption{
  Left: Comparison of the $E_p$ parameter between the \band and \issm functions. The dashed line is the equality line. The inset shows the ratio between the $E_p$ values obtained by \issm and \band functions. Right: limit energies
  \fpc{$E_\mathrm{lim}$} of the \issm and \band functions compared with their peak energies $E_p$. The \fpc{horizontal} dashed line represents the lower limit of the GBM energy range. 
  }
  \label{fig:Epeak_scatter}
\end{figure*}

In this section we compare the \fpc{spectral} parameters of the \band and \issm functions.
The left panel of Fig.~\ref{fig:Epeak_scatter} shows \fpc{the SED peak energies} obtained with the two
models. 
\fpc{The $E_p$} values of the \issm function are \fpc{found to be systematically} larger than the values obtained with the \band function.
The low-energy index $\alpha$ is an asymptotic value that is rarely reached by the local photon index within the energy
range of any burst-observing instrument.
\fpc{For this reason}, \cite{Preece1998} defined an effective low-energy index at the \fpc{CGRO/BATSE} detector lower limit ($25$
keV). 
In order to find the energy limit ($E_\mathrm{lim}$) at which the local photon index $\Gamma(E)$ approaches the asymptotic
value $\alpha$ within its error $\delta\alpha$, we solved the equation $\Gamma(E_\mathrm{lim})=\alpha-\delta\alpha$ using
the definition of the local photon index of the \band and \issm functions in Eqs.~\ref{eq:band_slope} and
\ref{eq:issm_slope}, respectively. The $E_\mathrm{lim}$ energies of the two functions are expressed as:
  \begin{equation}
E_\mathrm{lim, \band} = \frac{\delta\alpha}{2+\alpha}\,E_p
\label{eq:Elim_Band}
\end{equation}
\begin{equation}
E_\mathrm{lim, \issm} = \frac{\delta\alpha\,(2+\beta)}{(2+\alpha)(\beta - \alpha + \delta\alpha)}\,E_p
\label{eq:Elim_ISSM}
\end{equation}

\begin{figure*}[!t]
  \centering
  \hbox{
  \includegraphics[width=0.5\linewidth]{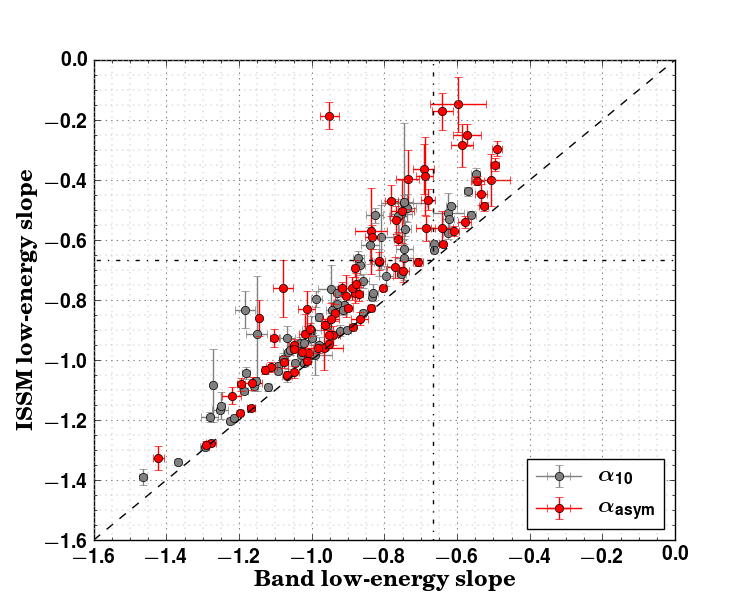}
  \includegraphics[width=0.5\linewidth]{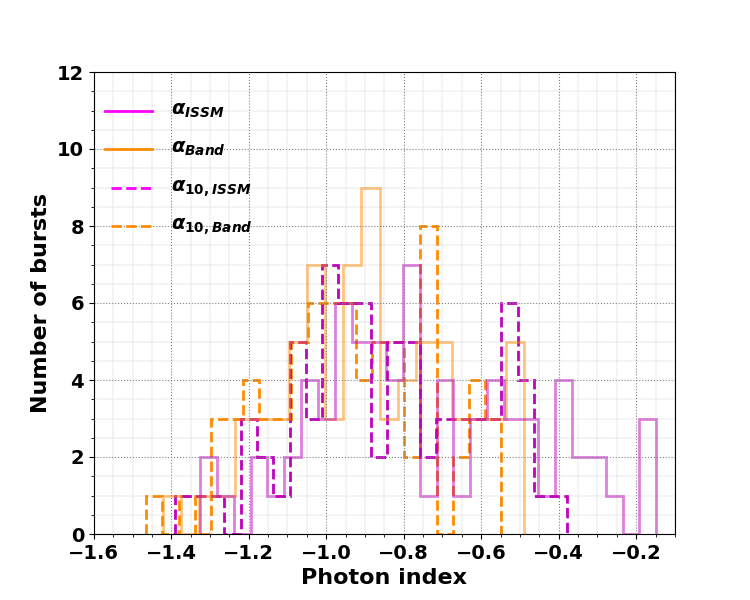}
  }
  \caption{
  Left: Comparison of the asymptotic $\alpha$ and the local photon index $\alpha_{10}$ at 10 keV between the \band
  and \issm functions. The gray dashed line denotes equality. The dashed-dotted horizontal and vertical lines indicate the upper limit ($-2/3$) of the low-energy spectral index for
  synchrotron emission \fpc{in the} slow cooling regime. Right: Distributions of the $\alpha$ and $\alpha_{10}$ parameters of the \band and \issm functions.
  }
  \label{fig:alpha_scatter}
\end{figure*}

These quantities are displayed with respect to $E_p$ in the right panel of Fig.~\ref{fig:Epeak_scatter}.
For the vast majority of the GRBs in our sample, the $E_\mathrm{lim}$ values fall below the GBM energy range.
We thus defined $\alpha_{10}$ as the local photon index at $10$ keV, namely right above the low-energy detection limit of the GBM.
The left panel of Fig.~\ref{fig:alpha_scatter} compares the $\alpha_{10}$ index to the $\alpha$ asymptotic index for both the \band and \issm functions.
While the $\alpha$ indices of the \issm function are larger than those of the \band function, the $\alpha_{10}$ indices
of the \issm function are only slightly larger.
The values of $\alpha_{10}$ appear also less scattered than those of $\alpha$.
More interestingly, the fraction of GRBs that are fitted with the \issm function and whose index is harder than the synchrotron slow-cooling limit ($-2/3$) decreases from
$35$\% ($\alpha$ asymptotic index) to $26$\% ($\alpha_{10}$).
This fraction decreases from $19$\% to $12$\% for the \band function.
As shown in the right panel of Fig.~\ref{fig:alpha_scatter} that displays the $\alpha$ and $\alpha_{10}$ distributions for
both models, the weighted mean index of the \issm (resp. \band) function indeed decreased from
$\langle\alpha\rangle=-0.75$ (resp. $-0.88$) to $\langle\alpha_{10}\rangle=-0.97$ (resp. $-1.03$).\\

Similarly, the $\beta$ parameter of the \issm function is an asymptotic value at high energy, which may not be
reached by the local photon index within the GBM energy range.
\fpc{Therefore}, we defined $\beta_b$ as the photon index at the break energy $E_b$
of the \band function \fpc{(Eq.~\ref{eq:band})}.
  By definition, $\beta_b$ is equal to $\beta$ for the \band function, while is it harder than $\beta$ for the \issm
  function owing to its continuous curvature.
  The $\beta_b$ index of the \issm function was also found to be systematically harder than that of the \band function, namely $\beta_\issm$ < $\beta_\band$ < $\beta_{b,\issm}$.
As a result, GRB spectra appear slightly wider around their peak energy when fitted with the \issm function than with the
\band function, but narrower when observed over a wider energy range.
This is illustrated in the right panel of Fig.~\ref{fig:asymetric_gaussian_fit}, for the case of GRB\,150403913,
\fpc{which is best fitted by the \issm model.}

\subsection{\band and \issm spectral sharpness}
\label{subsec:sharpness}
\begin{figure*}[!t]
  \centering
  \hbox{
  \includegraphics[width=0.5\linewidth]{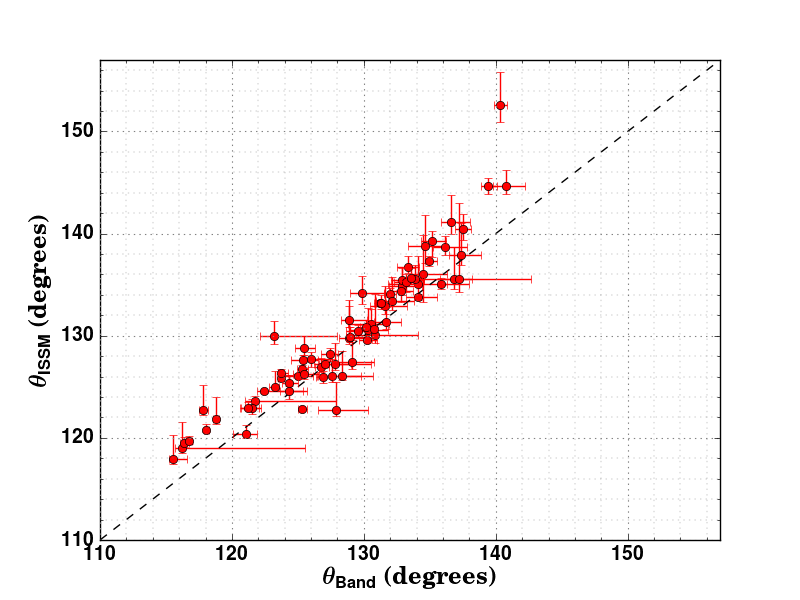}
  \includegraphics[width=0.5\linewidth]{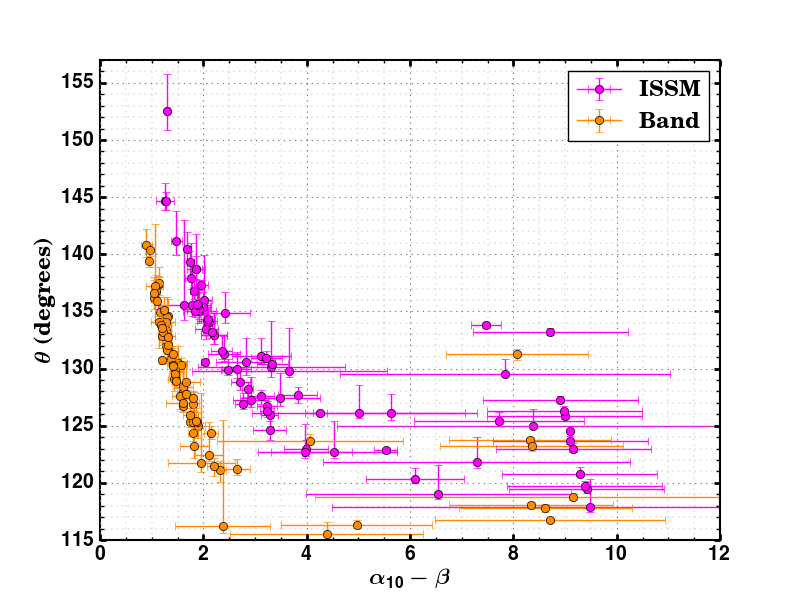}
   }
  \caption{
    Left: Spectral sharpness angles of the \issm fits versus the angles of the \band fits to the GRB spectra.
    Right: 
    Spectral sharpness angle as function of the difference between the $\alpha_{10}$ and $\beta$ parameters for the \band and \issm functions.
  }
  \label{fig:theta_vs_alpha-beta_Ep}
\end{figure*}
\begin{figure*}[t!]
  \centering
  \includegraphics[width=0.53\linewidth]{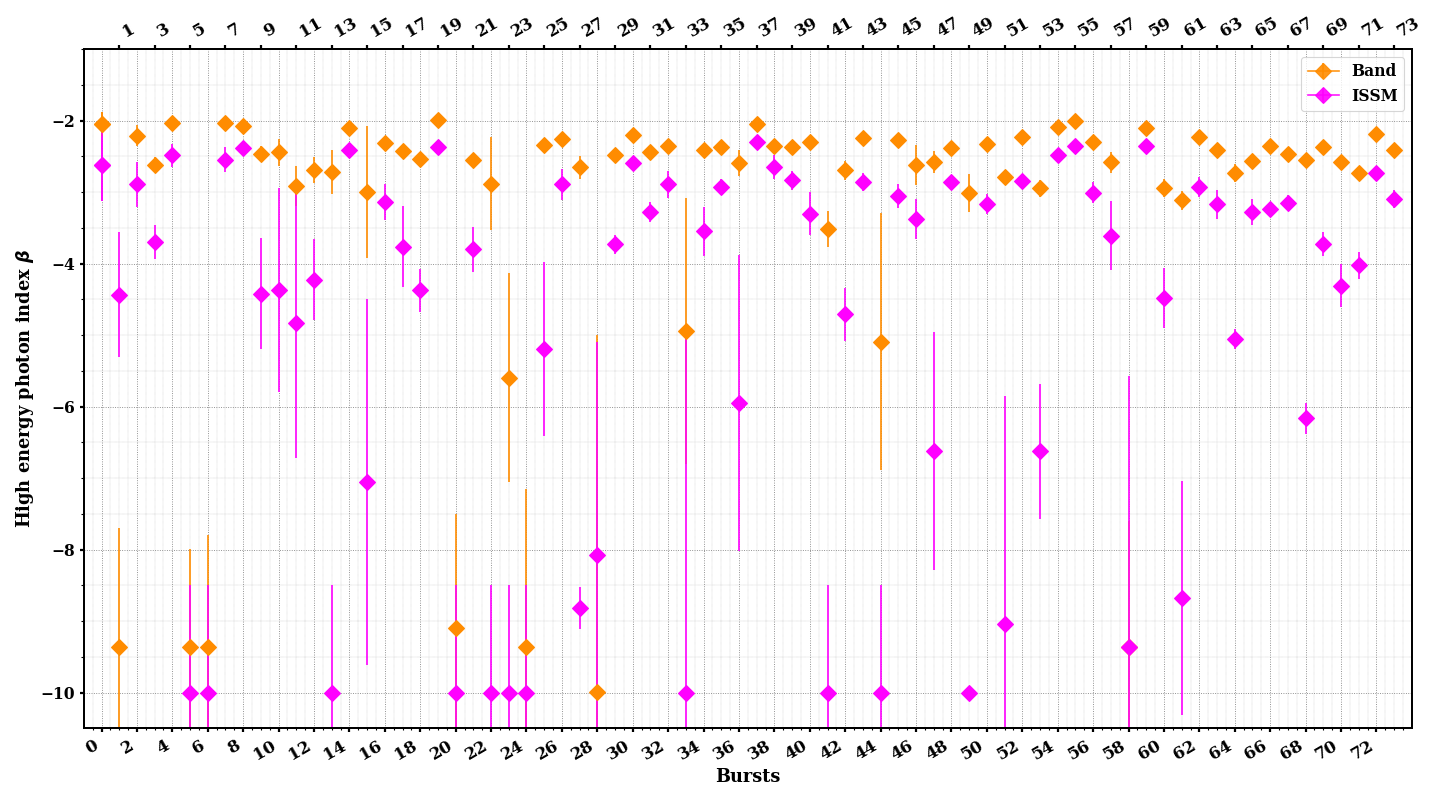}
  \includegraphics[width=0.40\linewidth]{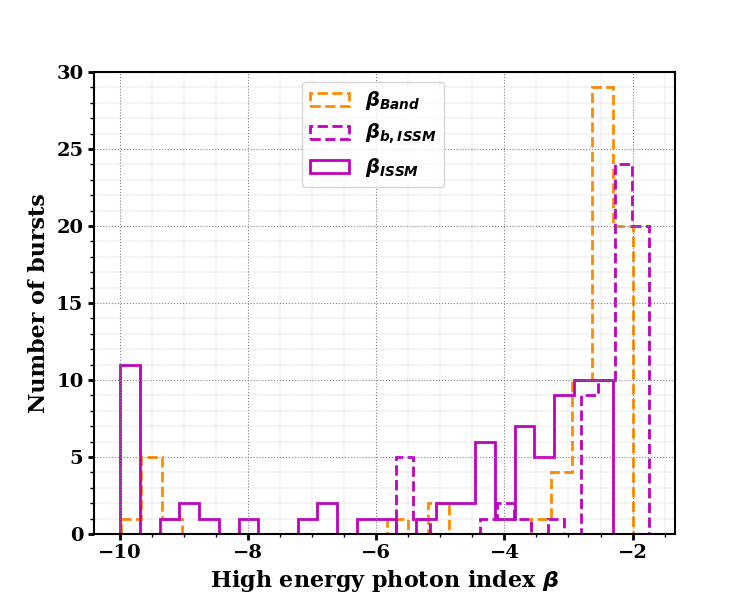}
  \caption{
  Left: Distribution of the high-energy index $\beta$ of the \band and \issm functions. The lowest limit value used for $\beta$ was fixed to -10. Right: Distributions of the $\beta$ parameter of the \band and \issm functions.
  }
  \label{fig:beta}
\end{figure*}
We investigated how the sharpness of the \band and \issm fitted spectra varies quantitatively with the photon indices. Following the methodology described in Sect.~\ref{sec:ana_sim}, a set of $10^3$ spectra was simulated for each GRB using its fit parameters and their covariance matrix.
The spectral sharpness angles of the GRB sample
are presented in the left panel of Fig.~\ref{fig:theta_vs_alpha-beta_Ep}.
Similarly to the synthetic bursts analyzed in Sect.~\ref{sec:ana_sim}, the \issm spectra are slightly wider
  than the \band spectra.
  As expected, the spectral sharpness angle was found to be independent of the peak energy, and to depend strongly on the photon indices.
As shown in the right panel of Fig.~\ref{fig:theta_vs_alpha-beta_Ep}, the \fpc{spectral sharpness} angle decreases with increasing $\alpha_{10}$ and/or with \fpc{decreasing} $\beta$.
The spectral sharpness angles of the GRBs in our sample are similar to those obtained by \cite{Yu2015A&A} (see figures
  therein, e.g., the blue solid curve in the left panel of Fig.~\ref{fig:Epeak_scatter}), ranging from $\sim
115^{\circ}$ to $\sim 140^{\circ}$ in both analyses, except one with angle 152$^{\circ}$ with the \issm function.

More importantly, the spectral sharpness angle of the synthetic bursts fitted by the \issm function is $149^{\circ}$ (see Table~\ref{table:angles_2bursts}), which is larger than for any GRB in our sample using the same fitting model.
This results essentially from the difference in the low-energy spectral index, $\alpha\simeq
-1.2$, which is softer than for most of the analyzed GRBs (see the left panel of Fig.~\ref{fig:alpha_scatter}). 
Besides, the value of the high-energy index of the synthetic burst, $\beta\simeq -2.3$, is close to the higher bound of the sample distribution as shown in Fig.~\ref{fig:beta}.
Possible ways to improve the agreement between the synthetic and observed bursts will be discussed in Sect.~\ref{sec:discussion}.


%
\section{Application to GRB\,090926A}
\label{sec:ana_090926A}
%
The prompt light curve of GRB\,090926A shows a short and bright spike at $10$ s post-trigger which has been detected from
\fpc{keV to GeV energies by the \Fermi instruments} \citep{Ackermann2011ApJ...729..114A}.
This spike coincides with the emergence of a hard power-law spectral component which is attenuated at the highest energies.
In Y17, we performed a dedicated analysis of the broadband prompt emission spectrum of GRB\,090926A by
combining the GBM data with the LAT Pass 8 data above $30$ MeV.
Using a \band+\cutbpl fitting function, we showed that the spectral break energy increases with time, and that the
entire prompt emission of this burst, namely the emission that is observed from keV to GeV energies by the GBM and the LAT
during the GRB duration in the 50-300 keV energy band, can be interpreted as the result of synchrotron emission of shock-accelerated electrons in the keV-MeV domain,
with an inverse Compton spectral component at higher energies.
The latter component was fitted by the \cutbpl function instead of the \cutpl function to
avoid any unrealistic contribution to the observed flux in the GBM low-energy range.
As a result, the low-energy index $\alpha$ of the \band spectral component was found to be close to $-0.9$, which is in
agreement with the theoretical index ($\sim -1$) of the fast-cooling synchrotron spectrum that is expected in the presence
of inverse Compton scatterings in the Klein-Nishina regime (BD14).\\

\begin{figure*}[t!]
  \centering
  \hbox{
  \includegraphics[width=0.5\linewidth]{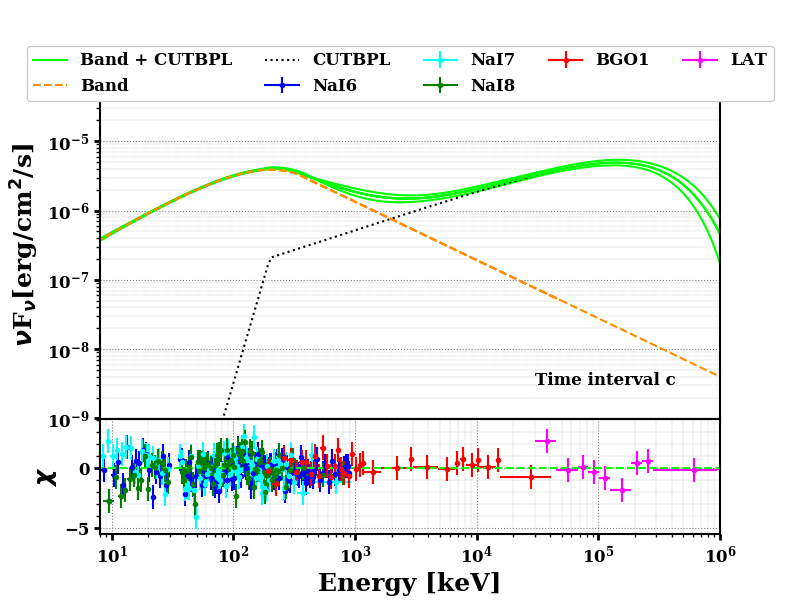}
  \includegraphics[width=0.5\linewidth]{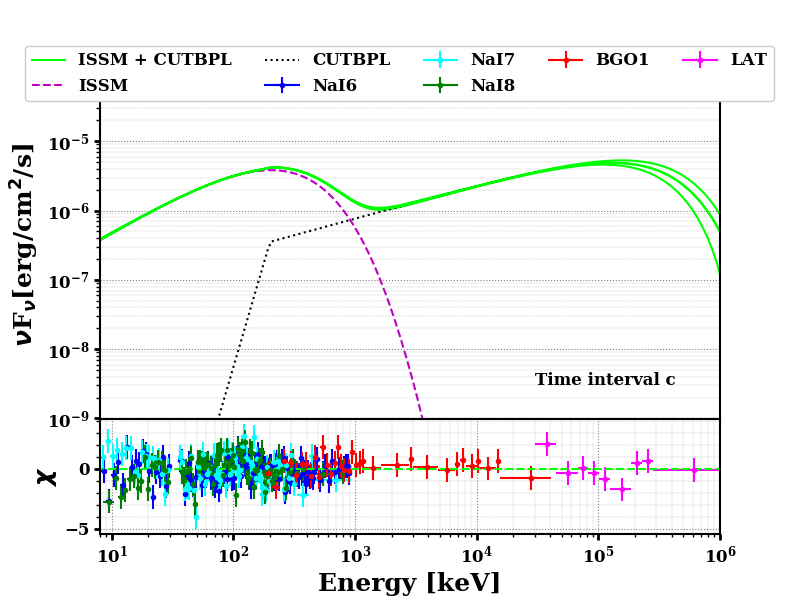}
   }
  \hbox{
  \includegraphics[width=0.5\linewidth]{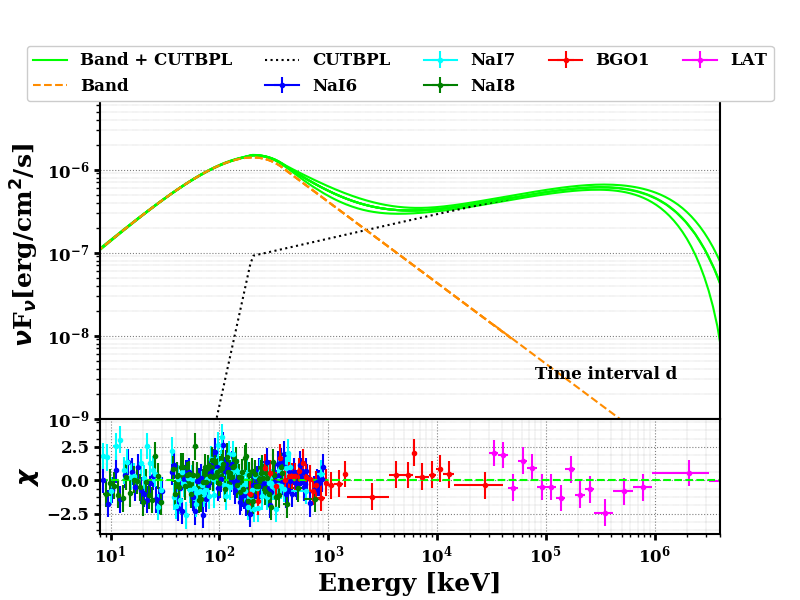}
  \includegraphics[width=0.5\linewidth]{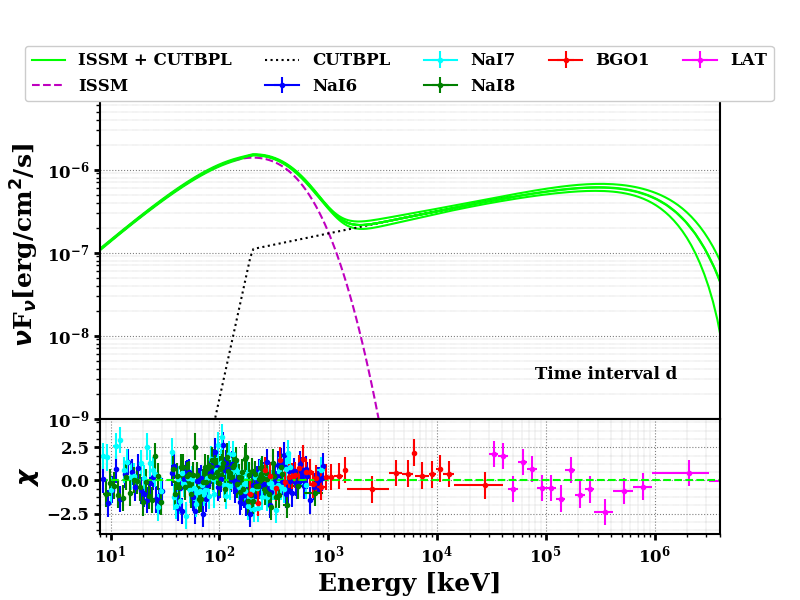}
   }
  \caption{
  Spectral energy distributions of GRB\,090926A in time intervals $c$ (top panels) and $d$ (lower panels) from the joint GBM/LAT analysis using LAT Pass 8 above $30$ MeV.
  The solid curve represents the \band + \cutbpl model \fpc{(left colum)} and the \issm + \cutbpl model \fpc{(right column)}, within a 68$\%$ confidence level contour derived from the errors on the fit parameters. The reference energy of the \cutbpl was fixed to $10$ MeV \fpc{like} in Y17.
  }
  \label{fig:090926A_d_IAP}
\end{figure*}

\fpc{Going further, we} revisited the spectral analysis of GRB\,090926A \fpc{and compared the \band + \cutbpl model to the \issm + \cutbpl model.}
This analysis was performed with the \fpc{\textit{XSPEC}} software for the time intervals \fpc{$c$ ($0.98$ s to $10.5$ s)
  and $d$ ($10.5$ s to $21.5$ s) where the high-energy break is detected.}
Following Y17, we fixed the parameters $\gamma_0$ and the break energy $E_b$ of Eq.~\ref{eq:cutbpl} to $+4$ and $200$ keV, respectively.
\fpc{Like} in Y17, the reference energy $E_0$ 
was fixed to $10$ MeV and $100$ MeV for the time intervals $c$ and $d$, respectively.
The results of this analysis are presented in Fig.~\ref{fig:090926A_d_IAP} and summarized in Table
\ref{table:IAP_fit_flux_keV_GeV_cutpl_cutbpl_090926A}.
\fpc{As can be seen in all panels of this figure and from the \pgstat fit statistics, both the \band+\cutbpl and \issm+\cutbpl models
  reproduce adequately the GRB spectrum, especially in the time interval $c$ (top panels).
The low-energy indices $\alpha$ of the \band and \issm spectral components are equal within statistical errors, and close
to $-1$ and $-0.9$ for the time intervals $c$ and $d$, respectively.
Again, these values perfectly agree with the predictions of BD14.}
\fpc{All other spectral parameters are also equivalent between both models,
  except the high-energy index $\beta$ of the keV-MeV spectral component, which is not well constrained using the \issm
  function. Since the \issm flux decreases more rapidly than that of the \band function beyond the SED peak energy, this likely results from the
  lack of photon statistics in the SED dip at a few MeV, between the GBM and LAT energy domains.
}


\bgroup
\def\arraystretch{1.5}%
\begin{table}[t!]
\centering
\caption{Results of the \band+\cutpl and \issm+\cutbpl fits to GBM/LAT data in the \fpc{time} intervals $c$ and $d$ of GRB\,090926A.}
\begin{tabular}{llcc}
 \hline
 Time intervals  &  Parameters  &   \band + \cutbpl &  \issm + \cutbpl\\
 \hline
[0.98 s -10.5 s] 
 &    $A_{\band/\issm}$ (\unit)  & $0.33_{-0.01}^{+0.02}$  & $580_{-5}^{+5}$($\times 10^{-5}$)   \\
 &    $E_p$ (keV)  & $203_{-7}^{+7}$  & $203_{-1}^{+1}$    \\
 &    $\alpha$  &   $-0.98_{-0.03}^{+0.03}$  &   $-0.97_{-0.02}^{+0.02}$    \\
 &    $\beta$& $-2.8_{-0.3}^{+0.2}$  & $-14.0_{-1.0}^{+4.7}$    \\
 &    $A_\cutbpl$ ($\times 10^4$ \unit)  &      $2.04_{-0.89}^{+1.37}$  &  $3.46_{-0.32}^{+0.22}$          \\
 &    $\lambda$   & $-1.43_{-0.11}^{+0.10}$  &  $-1.52_{-0.02}^{+0.02}$    \\
 &    $E_f$ (GeV) & $0.25_{-0.05}^{+0.07}$  & $0.27_{-0.05}^{+0.07}$    \\
 &    \pgstat/DOF &  577/510  &   582/510  \\
 \hline
[10.5 s - 21.5 s]
 &   $A_{\band/\issm}$ (\unit) & $0.123_{-0.003}^{+0.003}$  &        $203_{-11}^{+16}$($\times 10^{-5}$)    \\
 &   $E_p$ (keV)  & $201_{-5}^{+5}$  & $204_{-5}^{+7}$    \\
 &   $\alpha$  &   $-0.88_{-0.02}^{+0.02}$  &   $-0.86_{-0.02}^{+0.02}$    \\
 &   $\beta$ & $-3.0_{-0.3}^{+0.1}$  & $-14_{-1}^{+9}$    \\
 &   $A_\cutbpl$ ($\times 10^7$ \unit)  &    $8.89_{-2.72}^{+2.92}$  &    $10.74_{-2.93}^{+1.65}$       \\
 &   $\lambda$   & $-1.70_{-0.05}^{+0.06}$  &  $-1.73_{-0.04}^{+0.05}$    \\
 &   $E_f$ (GeV)  & $1.08_{-0.25}^{+0.31}$  & $1.12_{-0.25}^{+0.32}$    \\
 &   \pgstat/DOF &  714/510  &  717/510   \\
 \hline     
\end{tabular}
\label{table:IAP_fit_flux_keV_GeV_cutpl_cutbpl_090926A}
\end{table}


\section{Discussion}
\label{sec:discussion}


Our analysis of a sample of $74$ GRBs that are bright and fluent in the GBM showed that the
\issm function adequately reproduces most ($81$\%) of the \fpc{keV-MeV prompt emission} spectra, while the  \band
phenomenological function is suitable for a smaller fraction ($59$\%).
We observed noticeable differences between the spectra fitted with these two functions.
The peak energies $E_p$ of the spectra that are reconstructed using the \issm function are somewhat higher than those of the spectra
resulting from the \band function fits.
In addition, the \issm fitted spectra are globally narrower than the \band fitted spectra, yet they appear slightly wider
close to $E_p$.
This results in slightly larger sharpness angles for the \issm fitted spectra, which was also observed from fits of the synthetic spectrum.\\


\red{
Although the shape of the \issm function seems adequate to reproduce the spectral curvature of the GBM bright bursts, the spectral sharpness angle in this sample is always smaller than the sharpness angle of the synthetic spectra that were used to build this fitting function.
Since the spectral sharpness angle scales almost linearly with $\beta-\alpha$ (right panel of Fig.~\ref{fig:theta_vs_alpha-beta_Ep}), it is worth investigating possible ways to improve the agreement between the data and the physical model.
Firstly, the high-energy photon index $\beta$ in the model is strongly related to the slope of the electron power-law energy distribution $p$, as $\beta=-(p/2+1)$ in the synchrotron fast-cooling regime.
While $p=2.7$ and thus $\beta=-2.3$ for the synthetic bursts, larger values of $p$ up to $2.9$ could be
considered (BD14), owing to the theoretical uncertainties on the energy distribution of accelerated electrons in midly-relativistic shocks.
However, the expected change in the value of $\beta$ would not entirely account for the sharpness discrepancy with observed spectra.
Moreover, many of the observed values of $\beta$ are larger for the bursts with well-measured spectral parameters as shown in Fig.~\ref{fig:beta}. In this figure, softer high-energy indices appear with larger uncertainties and might be underestimated due to insufficient photon statistics above the peak energy.
This suggests that a better spectral coverage at MeV energies could result in harder values of $\beta$ and in larger sharpness angles for this fraction of the burst sample, making the entire sample compatible with the physical model.\\
}

Secondly, the low-energy photon index $\alpha$ of the synthetic spectra is close to $-1.2$. As shown in Fig.~\ref{fig:theta_vs_alpha-beta_Ep}, an increase of $0.5$ in this theoretical slope, with a condition that the high-energy slope does not increase, would be enough to make the synthetic spectra compatible with the GBM sample in terms of spectral sharpness.
As a matter of fact, harder values of $\alpha$ are expected from internal shock synchrotron models in the so-called
marginally fast-cooling regime \citep{DB11,Beniamini2013}, where the impact of adiabatic losses on the electron
energy distribution is not negligible anymore as compared to the effect of radiative losses. 
In this regime, specific configurations of the jet such as low-contrast internal shocks can lead to $\alpha$ values as
hard as $-2/3$ \citep{DB11}.
Hardening the low-energy part of the synchrotron spectrum could also be obtained by accounting for the decay of the magnetic field behind the shock \citep{peerzhang,derishev07}.
In such a configuration, the most energetic electrons would indeed explore a small region where the magnetic field has not decreased yet, while the less energetic electrons would see a less intense magnetic field on average.
Therefore, such a magnetic field decay appears as a natural possibility to reach the marginally fast cooling regime without any need for a fine-tuning of the microphysical parameters (Daigne \& Bo{\v s}njak, in preparation). Indeed, as the magnetic field decreases, the critical Lorentz factor of electrons for slow cooling $\gamma_\mathrm{c}$ increases. 

Preliminary results show that a steep asymptotic slope $\alpha$ close to $-2/3$ is obtained. This effect of a magnetic
field decay is shown in the right panel of Fig.~\ref{fig:slopes_function_fit} where the value of the \issm low-energy
slope $\alpha= b/c-a$ varies between $-1.5$ (case A with a constant magnetic field) and $-0.8$ (case B  with a magnetic
field decay for the spectrum measured at the peak of the light curve). This illustrates that a more realistic modeling of
the microphysics in the acceleration and emission regions should be investigated to reach a full agreement between the
synthetic and observed spectra.
Ultimately one would like to use the spectral fits to infer the physical parameters of the model
    such as the evolution of the injected power or the distribution of the Lorentz factor in the flow. In practice,
    analysis of the spectra only provides values for the four parameters: $A_\issm$, $\alpha$, $\beta$ and $E_p$. They
    partially constrain the shock physics and radiative mechanism as discussed above for $\alpha$ and $\beta$. The peak
    energy $E_p$ depends on a combination of the ejecta physical parameters and shock microphysics. It will therefore be
    difficult to decipher from the evolution of $E_p$ the form of the Lorentz factor or/and injected power distributions
    even if some general trends can probably be obtained. This will require a dedicated study.


\section{Conclusions}
\label{sec:concl}
\red{
The physical origin of GRB prompt emission remains elusive despite decades of observations.
Characterizing the prompt emission spectra has been often performed using phenomenological parameterizations with little physical
grounds, such as the \band function. 
However, the advance of instrument spectral coverage and the improved data quality provided by current missions such as
the \Fermi observatory now offer the possibility to confront observations to theoretical models in detail.
In this work, we used the internal shock model developed by BD14 to produce synthetic GRBs (see also \cite{BD09} and \cite{DB11}), and we folded their spectra with the response of the \Fermi GBM and LAT.
The synthetic spectra obtained from these simulations in the keV-MeV domain, where the synchrotron emission is dominant,
were used to build a new GRB spectral fitting function called \issm, which has the same number of parameters as the \band function.
We used the \issm function to fit the prompt emission spectra for a sample of $74$ GBM fluent bursts, which improved
the fit quality as compared to the phenomenological \band function in a sizeable number of cases.
In addition, we combined the \issm function with a \cutbpl spectral component to fit the GRB\,090926A
broadband spectrum with some success. This work was motivated by a previous study of this burst that suggested an internal origin of the keV to GeV emission observed during the prompt phase (Y17).
In this framework, our interpretation of both spectral components as being from synchrotron and inverse Compton emissions
would greatly benefit from a more realistic parameterization of the high-energy component based on the synthetic spectra,
especially in the overlapping region at MeV energies.\\
} 

\red{
The analysis of the GBM sample of $74$ bursts showed noticeable differences between the \issm and \band fits.
Peak energies and spectral sharpness angles that are obtained from the \issm fits are slightly larger than those from the \band
fits. This result can be attributed to the continuous curvature of the \issm function. This curvature reflects the time
evolution of the electron and photon energy distributions within the analysed time intervals, which lasts longer than the
typical dynamical timescales in the physical model.
While observed spectra can be well fitted by the \issm physical function, they appear narrower than the synthetic spectra,
essentially because of a theoretical low-energy photon index that differs significantly from the observed photon index
$\alpha$. 
This problem clearly calls for improvements of the internal shock model and possible solutions have been identified.
In particular, more sophisticated prescriptions for the jet physics should be investigated in the future, such as the
marginally fast-cooling regime and the decay of the magnetic field behind the shocks. Inferring
    the parameters of the physical model from the fitted parameters of the \issm function is not easy as their relation is
    complex. Actually, the physical parameters that best reproduce GRB prompt emission spectra should be rather explored
    by fitting the numerical model directly to the data in the future, without using the \issm proxy function. On
    the experimental side, complementary multi-wavelength observations will be also performed by GRB-dedicated missions
such as SVOM which will observe the complete time evolution of GRBs from possible precursors until the afterglow
phase~\citep{wei2016}. 
SVOM will measure GRB prompt emission spectra down to $4$ keV thanks to its ECLAIRs coded-mask telescope, and up to the MeV range with its Gamma-Ray Monitor detector~\citep{bernardini2017}. This will provide more insight into the
physical origin of GRB high-energy emission at early times.
}

\begin{acknowledgements}
The \Fermi LAT Collaboration acknowledges generous ongoing support
from a number of agencies and institutes that have supported both the
development and the operation of the LAT as well as scientific data analysis.
These include the National Aeronautics and Space Administration and the
Department of Energy in the United States, the Commissariat \`a l'Energie Atomique
and the Centre National de la Recherche Scientifique / Institut National de Physique
Nucl\'eaire et de Physique des Particules in France, the Agenzia Spaziale Italiana
and the Istituto Nazionale di Fisica Nucleare in Italy, the Ministry of Education,
Culture, Sports, Science and Technology (MEXT), High Energy Accelerator Research
Organization (KEK) and Japan Aerospace Exploration Agency (JAXA) in Japan, and
the K.~A.~Wallenberg Foundation, the Swedish Research Council and the
Swedish National Space Board in Sweden.
 
Additional support for science analysis during the operations phase is gratefully acknowledged from the Istituto Nazionale
di Astrofisica in Italy and the Centre National d'\'Etudes Spatiales in France.

\end{acknowledgements}
\begin{appendix}


\section{Spectral analysis results}
\label{sec:results}
\begin{table*}[h!]
  \centering
  \caption{
    Results of the  \band and  \issm fits for $\GRBBone$ in the four time intervals.
    \fpc{The true peak energies are $E_{p, true}= 1150$, $478$, $114$ and $745$} keV for the time intervals [0 s, 1 s], [1 s, 3 s], [3 s, 6 s] and [0 s, 6 s] respectively.
  }%
  \begin{tabular}{lcccccc}
\hline
 \hline Time interval  &  Model   &    $E_p$    & $E_p$/$E_{p, true}$ &  $\alpha$   &  $\beta$   &   Amplitude  \\
    &     &      &     &     &   ($\times 10^{-3}$ cm$^{-2}$\,s$^{-1}$\,keV$^{-1}$) \\
 \hline [0 s, 1 s]    &   \band     &  761 $\pm$ 14     & 0.66 &  -1.14 $\pm$ 0.01     &    -2.16 $\pm$ 0.01    &    1962 $\pm$ 10\\
              &    \issm    &    1216 $\pm$ 25    & 1.06  & -1.07 $\pm$ 0.01     &    -2.45 $\pm$ 0.02     &    36 $\pm$ 1\\
 \hline [1 s, 3 s]    &   \band     &   295 $\pm$ 3    & 0.62 &  -1.22 $\pm$ 0.01     &    -2.13 $\pm$ 0.01     &    3100 $\pm$ 19\\
               &    \issm     &    459 $\pm$ 6    & 0.96  & -1.09 $\pm$ 0.01     &    -2.35 $\pm$ 0.01     &    145 $\pm$ 1\\
 \hline [3 s, 6 s]    &   \band     &   99 $\pm$ 3    &  0.87 & -1.37 $\pm$ 0.02     &    -2.16 $\pm$ 0.01     &    477 $\pm$ 15\\
               &    \issm     &    119 $\pm$ 3    & 1.04  & -0.95 $\pm$ 0.08    &    -2.29 $\pm$ 0.02    &    192 $\pm$ 1\\
 \hline [0 s, 6 s]    &   \band     &   378 $\pm$ 4  & 0.51 &    -1.27 $\pm$ 0.01     &    -2.09 $\pm$ 0.01     &    1465 $\pm$ 6\\
               &    \issm     &    659 $\pm$ 9    & 0.88  & -1.16 $\pm$ 0.01     &    -2.28 $\pm$ 0.01    &    34 $\pm$ 1\\
\hline
  \end{tabular}
  \label{table:Band_fit_flux}
\end{table*}
\begin{table*}[h!]
  \centering
  \caption{
  Results of the  \band and  \issm fits for $\GRBBten$ in the four time intervals.
    \fpc{The true peak energies are $E_{p, true}= 1150$, $478$, $114$ and $745$} keV for the time intervals [0 s, 1 s], [1 s, 3 s], [3 s, 6 s] and [0 s, 6 s] respectively.
  }%
  \begin{tabular}{lcccccc}
 \hline
 \hline Time interval  &  Model   &    $E_p$    & $E_p$/$E_{p, true}$  &  $\alpha$   &  $\beta$   &   Amplitude  \\
    &     &      &     &     &   ($\times 10^{-3}$ cm$^{-2}$\,s$^{-1}$\,keV$^{-1}$) \\
 \hline   [0 s, 1 s]  &   \band     &  677 $\pm$ 40  &  0.59 &    -1.13 $\pm$ 0.01     &    -2.11 $\pm$ 0.04     &    2008 $\pm$ 37\\
               &    \issm   &    1178 $\pm$ 82    & 1.02 &  -1.06 $\pm$ 0.02     &    -2.41 $\pm$ 0.07     &    35 $\pm$ 1\\
 \hline   [1 s, 3 s]  &   \band     &   298 $\pm$ 11  & 0.62  &    -1.23 $\pm$ 0.01     &    -2.13 $\pm$ 0.02     &    3077 $\pm$ 58\\
               &    \issm     &    470 $\pm$ 18    & 0.98 &   -1.08 $\pm$ 0.03     &    -2.33 $\pm$ 0.03    &    145 $\pm$ 2\\
 \hline   [3 s, 6 s]  &   \band     &   104 $\pm$ 11 &  0.91  &    -1.44 $\pm$ 0.05    &    -2.13 $\pm$ 0.03     &    424 $\pm$ 41\\
               &    \issm     &    127 $\pm$ 11     & 1.11  & -1.02 $\pm$ 0.27     &    -2.23 $\pm$ 0.05     &    187 $\pm$ 3\\
 \hline   [0 s, 6 s]  &   \band     &   377 $\pm$ 13 &  0.51  &    -1.27 $\pm$ 0.01     &    -2.07 $\pm$ 0.02    &    1456 $\pm$ 20\\
               &    \issm    &    685 $\pm$ 32     & 0.92  & -1.16 $\pm$ 0.02     &    -2.26 $\pm$ 0.03     &    34 $\pm$ 1\\
\hline
  \end{tabular}
  \label{table:Band_fit_flux10}
\end{table*}
\begin{table*}[h!]
  \centering
  \caption{
  Results of the  \band and  \issm fits for $\GRBBhun$ in the four time intervals.  
    \fpc{The true peak energies are $E_{p, true}= 1150$, $478$, $114$ and $745$} keV for the time intervals [0 s, 1 s], [1 s, 3 s], [3 s, 6 s] and [0 s, 6 s] respectively.
  }%
  \begin{tabular}{lcccccc}
 \hline
 \hline Time interval  &  Model   &    $E_p$   & $E_p$/$E_{p, true}$   &  $\alpha$   &  $\beta$   &   Amplitude  \\
    &     &      &     &     &   ($\times 10^{-3}$ cm$^{-2}$\,s$^{-1}$\,keV$^{-1}$) \\
 \hline   [0 s, 1 s]  &   \band     &  709 $\pm$ 135 & 0.62   &    -1.09 $\pm$ 0.05     &    -2.07 $\pm$ 0.12     &    193 $\pm$ 11\\
                       &    \issm   &    1333 $\pm$ 324  & 1.16  &    -1.03 $\pm$ 0.08     &    -2.40 $\pm$ 0.24     &    4 $\pm$ 1\\
 \hline   [1 s, 3 s]  &   \band     &   297 $\pm$ 37 &  0.62  &    -1.25 $\pm$ 0.04     &    -2.13 $\pm$ 0.07     &    301 $\pm$ 18\\
                       &    \issm     &    457 $\pm$ 50  &  0.96 &    -1.18 $\pm$ 0.07     &    -2.45 $\pm$ 0.16     &    14 $\pm$ 1\\
 \hline   [3 s, 6 s]  &   \band     &   55 $\pm$ 17  & 0.48  &    -0.90 $\pm$ 0.38     &    -2.00 $\pm$ 0.06     &    136 $\pm$ 121\\
                       &    \issm     &    130 $\pm$ 34  & 1.14  &    unconstrained     &    -2.18 $\pm$ 0.13     &    19 $\pm$ 1\\
 \hline   [0 s, 6 s]  &   \band     &   445 $\pm$ 50  &  0.60 &    -1.31 $\pm$ 0.03     &    -2.13 $\pm$ 0.07     &    137 $\pm$ 5\\
                       &    \issm     &    678 $\pm$ 78  & 0.91  &    -1.24 $\pm$ 0.04     &    -2.38  $\pm$ 0.12     &    3.5 $\pm$ 0.1\\
\hline
  \end{tabular}
  \label{table:Band_fit_flux100}
\end{table*}
\newpage
\footnotesize
\setlength\LTleft{0pt}
\setlength\LTright{0pt}
\begin{longtable}{@{\extracolsep{\fill}}ccccccccr@{}}
  \caption{Results of the \band and \issm spectral fits to GBM data for the prompt emission of \fpc{74 GRBs. The amplitudes are given in units of $10^{-4}\,$cm$^{-2}$\,s$^{-1}$\,keV$^{-1}$.}}\\
\hline
 GRB name  &  Models   &     $E_p$     &  $\alpha$   & $\alpha_{10}$ &  $\beta$   &  $\beta_b$ &   Amplitude  & \pgstat/DOF\\
 \hline
 \endfirsthead
\multicolumn{9}{c}{\tablename\ \thetable\ -- \textit{Continued from previous page}} \\
\hline
 GRB name  &  Models   &     $E_p$     &  $\alpha$   & $\alpha_{10}$ &  $\beta$   &  $\beta_b$ &   Amplitude  & \pgstat/DOF\\
 \hline
 \endhead
\hline \multicolumn{9}{r}{\textit{Continued on next page}} \\
\endfoot
\endlastfoot
\hline
GRB080817161  &     \band   & 410  $\pm$    14  &   -0.96 $\pm$  0.01  &   -1.00  $\pm$  0.01  &   -2.32  $\pm$  0.08  &   -2.32  $\pm$  0.08  &    145  $\pm$     2  &   1031  /   469  \\                       &     \issm   & 509  $\pm$    11  &   -0.88  $\pm$  0.02  &   -0.93  $\pm$  0.01  &   -3.13  $\pm$  0.25  &   -2.03  $\pm$  0.02  &    8.9  $\pm$   0.2  &   1021  /   469  \\  
GRB080825593  &     \band   & 187  $\pm$     7  &   -0.64 $\pm$  0.03  &   -0.75  $\pm$  0.02  &   -2.35  $\pm$  0.10  &   -2.35  $\pm$  0.10  &    641  $\pm$    30  &   1144  /   469  \\                       &     \issm   & 211  $\pm$     5  &   -0.56  $\pm$  0.06  &   -0.66  $\pm$  0.04  &   -5.19  $\pm$  1.22  &   -2.11  $\pm$  0.03  &   10.6  $\pm$   0.2  &   1149  /   469  \\  
GRB081125496  &     \band   & 183  $\pm$     8  &   -0.51 $\pm$  0.05  &   -0.62  $\pm$  0.04  &   -3.00  $\pm$  0.92  &   -3.00  $\pm$  0.92  &    913  $\pm$    72  &   534  /   351  \\                       &     \issm   & 187  $\pm$     6  &   -0.40  $\pm$  0.09  &   -0.51  $\pm$  0.07  &   -7.06  $\pm$  2.56  &   -2.67  $\pm$  0.10  &   10.2  $\pm$   0.7  &   532  /   351  \\  
GRB081207680  &     \band   & 705  $\pm$    40  &   -0.77 $\pm$  0.02  &   -0.80  $\pm$  0.02  &   -2.62  $\pm$  0.28  &   -2.62  $\pm$  0.28  &     75  $\pm$     1  &   1794  /   353  \\                       &     \issm   & 868  $\pm$    39  &   -0.69  $\pm$  0.04  &   -0.72  $\pm$  0.03  &   -3.37  $\pm$  0.28  &   -2.14  $\pm$  0.04  &    8.6  $\pm$   0.2  &   1777  /   353  \\  
GRB081224887  &     \band   & 404  $\pm$    10  &   -0.71 $\pm$  0.01  &   -0.75  $\pm$  0.01  &   -9.09  $\pm$  1.58  &   -9.09  $\pm$  1.58  &    372  $\pm$     6  &   648  /   474  \\                       &     \issm   & 411  $\pm$     7  &   -0.67  $\pm$  0.01  &   -0.71  $\pm$  0.01  &   -10.00  $\pm$  1.50  &   -5.47  $\pm$  0.03  &   23.8  $\pm$   0.3  &   647  /   474  \\  
GRB090328401  &     \band   & 754  $\pm$    51  &   -1.05 $\pm$  0.02  &   -1.07  $\pm$  0.01  &   -2.44  $\pm$  0.19  &   -2.44  $\pm$  0.19  &     98  $\pm$     2  &   1241  /   473  \\                       &     \issm   & 897  $\pm$    80  &   -1.04  $\pm$  0.02  &   -1.05  $\pm$  0.02  &   -4.37  $\pm$  1.42  &   -2.15  $\pm$  0.04  &    9.6  $\pm$   0.2  &   1243  /   473  \\  
GRB090528516  &     \band   & 154  $\pm$     7  &   -0.84 $\pm$  0.04  &   -0.95  $\pm$  0.04  &   -2.04  $\pm$  0.05  &   -2.04  $\pm$  0.05  &    197  $\pm$    14  &   2652  /   472  \\                       &     \issm   & 241  $\pm$    24  &   -0.57  $\pm$  0.14  &   -0.76  $\pm$  0.08  &   -2.55  $\pm$  0.18  &   -1.82  $\pm$  0.04  &    3.9  $\pm$   0.2  &   2650  /   472  \\  
GRB090618353  &     \band   & 164  $\pm$     3  &   -1.10 $\pm$  0.01  &   -1.18  $\pm$  0.01  &   -2.46  $\pm$  0.04  &   -2.46  $\pm$  0.04  &    720  $\pm$    15  &   1229  /   238  \\                       &     \issm   & 171  $\pm$     2  &   -0.93  $\pm$  0.03  &   -1.04  $\pm$  0.02  &   -3.15  $\pm$  0.11  &   -2.21  $\pm$  0.01  &   13.0  $\pm$   0.2  &   1173  /   238  \\  
GRB090718762  &     \band   & 170  $\pm$     5  &   -1.11 $\pm$  0.01  &   -1.19  $\pm$  0.01  &   -2.69  $\pm$  0.18  &   -2.69  $\pm$  0.18  &    312  $\pm$     8  &   666  /   469  \\                       &     \issm   & 173  $\pm$     2  &   -1.02  $\pm$  0.02  &   -1.10  $\pm$  0.01  &   -4.22  $\pm$  0.57  &   -2.41  $\pm$  0.03  &    5.3  $\pm$   0.3  &   662  /   469  \\  
GRB090719063  &     \band   & 240  $\pm$     2  &   -0.54 $\pm$  0.02  &   -0.62  $\pm$  0.02  &   -2.95  $\pm$  0.12  &   -2.95  $\pm$  0.12  &   1281  $\pm$    30  &   460  /   354  \\                       &     \issm   & 250  $\pm$     4  &   -0.45  $\pm$  0.03  &   -0.53  $\pm$  0.03  &   -6.62  $\pm$  0.94  &   -2.59  $\pm$  0.03  &   30.9  $\pm$   0.6  &   455  /   354  \\  
GRB090809978  &     \band   & 175  $\pm$    10  &   -0.74 $\pm$  0.03  &   -0.84  $\pm$  0.02  &   -1.98  $\pm$  0.04  &   -1.98  $\pm$  0.04  &    677  $\pm$    35  &   815  /   471  \\                       &     \issm   & 344  $\pm$    46  &   -0.40  $\pm$  0.10  &   -0.62  $\pm$  0.06  &   -2.37  $\pm$  0.09  &   -1.75  $\pm$  0.02  &   16.7  $\pm$   0.5  &   810  /   471  \\  
GRB090829672  &     \band   & 196  $\pm$     9  &   -1.42 $\pm$  0.01  &   -1.46  $\pm$  0.01  &   -2.36  $\pm$  0.10  &   -2.36  $\pm$  0.10  &    280  $\pm$     8  &   510  /   237  \\                       &     \issm   & 208  $\pm$    10  &   -1.33  $\pm$  0.04  &   -1.39  $\pm$  0.03  &   -2.64  $\pm$  0.17  &   -2.14  $\pm$  0.02  &    8.0  $\pm$   0.2  &   498  /   237  \\  
GRB091003191  &     \band   & 397  $\pm$    16  &   -0.94 $\pm$  0.02  &   -0.98  $\pm$  0.02  &   -2.59  $\pm$  0.19  &   -2.59  $\pm$  0.19  &    272  $\pm$     7  &   551  /   355  \\                       &     \issm   & 429  $\pm$    19  &   -0.92  $\pm$  0.03  &   -0.95  $\pm$  0.03  &   -5.95  $\pm$  2.07  &   -2.34  $\pm$  0.06  &   16.1  $\pm$   0.5  &   552  /   355  \\  
GRB091120191  &     \band   & 136  $\pm$     5  &   -1.16 $\pm$  0.03  &   -1.25  $\pm$  0.02  &   -2.92  $\pm$  0.28  &   -2.92  $\pm$  0.28  &    193  $\pm$     9  &   965  /   470  \\                       &     \issm   & 134  $\pm$     4  &   -1.08  $\pm$  0.02  &   -1.17  $\pm$  0.01  &   -4.83  $\pm$  1.89  &   -2.61  $\pm$  0.04  &    2.1  $\pm$   0.3  &   964  /   470  \\  
GRB091128285  &     \band   & 192  $\pm$     1  &   -0.95 $\pm$  0.01  &   -1.03  $\pm$  0.01  &   -2.58  $\pm$  0.16  &   -2.58  $\pm$  0.16  &    160  $\pm$     1  &   1037  /   353  \\                       &     \issm   & 199  $\pm$     2  &   -0.92  $\pm$  0.02  &   -0.98  $\pm$  0.01  &   -6.62  $\pm$  1.66  &   -2.40  $\pm$  0.03  &    2.8  $\pm$   0.0  &   1041  /   353  \\  
GRB100322045  &     \band   & 333  $\pm$    10  &   -0.88 $\pm$  0.01  &   -0.93  $\pm$  0.01  &   -2.20  $\pm$  0.04  &   -2.20  $\pm$  0.04  &    307  $\pm$     6  &   779  /   469  \\                       &     \issm   & 487  $\pm$    23  &   -0.69  $\pm$  0.03  &   -0.78  $\pm$  0.02  &   -2.60  $\pm$  0.07  &   -1.91  $\pm$  0.02  &   15.6  $\pm$   0.2  &   726  /   469  \\  
GRB100324172  &     \band   & 461  $\pm$    12  &   -0.58 $\pm$  0.02  &   -0.62  $\pm$  0.02  &   -5.60  $\pm$  1.46  &   -5.60  $\pm$  1.46  &    369  $\pm$     6  &   627  /   469  \\                       &     \issm   & 468  $\pm$     9  &   -0.54  $\pm$  0.02  &   -0.58  $\pm$  0.02  &   -10.00  $\pm$  1.50  &   -4.21  $\pm$  0.04  &   30.5  $\pm$   0.5  &   631  /   469  \\  
GRB100414097  &     \band   & 637  $\pm$    12  &   -0.53 $\pm$  0.01  &   -0.56  $\pm$  0.01  &   -4.95  $\pm$  1.86  &   -4.95  $\pm$  1.86  &    349  $\pm$     4  &   1070  /   471  \\                       &     \issm   & 651  $\pm$    12  &   -0.49  $\pm$  0.01  &   -0.52  $\pm$  0.01  &   -10.00  $\pm$  5.00  &   -3.88  $\pm$  0.03  &   46.1  $\pm$   0.3  &   1090  /   471  \\  
GRB100511035  &     \band   & 625  $\pm$    38  &   -1.28 $\pm$  0.01  &   -1.29  $\pm$  0.01  &   -9.37  $\pm$  1.37  &   -9.37  $\pm$  1.37  &     94  $\pm$     2  &   798  /   473  \\                       &     \issm   & 656  $\pm$    51  &   -1.28  $\pm$  0.01  &   -1.29  $\pm$  0.01  &   -10.00  $\pm$  1.50  &   -5.56  $\pm$  0.02  &    6.8  $\pm$   0.1  &   798  /   473  \\  
GRB100612726  &     \band   & 113  $\pm$     2  &   -0.57 $\pm$  0.04  &   -0.75  $\pm$  0.03  &   -2.55  $\pm$  0.07  &   -2.55  $\pm$  0.07  &   1290  $\pm$    79  &   524  /   472  \\                       &     \issm   & 121  $\pm$     2  &   -0.25  $\pm$  0.04  &   -0.51  $\pm$  0.02  &   -3.80  $\pm$  0.32  &   -2.23  $\pm$  0.02  &    6.3  $\pm$   0.6  &   528  /   472  \\  
GRB100707032  &     \band   & 266  $\pm$    14  &   -0.69 $\pm$  0.03  &   -0.76  $\pm$  0.02  &   -2.08  $\pm$  0.05  &   -2.08  $\pm$  0.05  &    236  $\pm$    10  &   450  /   236  \\                       &     \issm   & 504  $\pm$    61  &   -0.36  $\pm$  0.08  &   -0.52  $\pm$  0.06  &   -2.39  $\pm$  0.10  &   -1.79  $\pm$  0.02  &    9.7  $\pm$   0.2  &   440  /   236  \\  
GRB100719989  &     \band   & 321  $\pm$    12  &   -0.69 $\pm$  0.03  &   -0.74  $\pm$  0.02  &   -2.41  $\pm$  0.08  &   -2.41  $\pm$  0.08  &    462  $\pm$    15  &   733  /   354  \\                       &     \issm   & 384  $\pm$    13  &   -0.56  $\pm$  0.04  &   -0.63  $\pm$  0.03  &   -3.55  $\pm$  0.34  &   -2.07  $\pm$  0.03  &   22.4  $\pm$   0.4  &   726  /   354  \\  
GRB100826957  &     \band   & 461  $\pm$    25  &   -1.05 $\pm$  0.01  &   -1.08  $\pm$  0.01  &   -2.05  $\pm$  0.02  &   -2.05  $\pm$  0.02  &    310  $\pm$     5  &   717  /   237  \\                       &     \issm   &1005  $\pm$    77  &   -0.95  $\pm$  0.02  &   -1.00  $\pm$  0.02  &   -2.30  $\pm$  0.04  &   -1.80  $\pm$  0.02  &   21.1  $\pm$   0.2  &   696  /   237  \\  
GRB100829876  &     \band   & 136  $\pm$     5  &   -0.60 $\pm$  0.08  &   -0.75  $\pm$  0.06  &   -2.04  $\pm$  0.04  &   -2.04  $\pm$  0.04  &    946  $\pm$   104  &   276  /   237  \\                       &     \issm   & 232  $\pm$    29  &   -0.15  $\pm$  0.09  &   -0.48  $\pm$  0.03  &   -2.49  $\pm$  0.16  &   -1.77  $\pm$  0.06  &   13.8  $\pm$   0.8  &   275  /   237  \\  
GRB100910818  &     \band   & 159  $\pm$    10  &   -0.94 $\pm$  0.01  &   -1.03  $\pm$  0.00  &   -2.46  $\pm$  0.11  &   -2.46  $\pm$  0.11  &    376  $\pm$     8  &   587  /   469  \\                       &     \issm   & 168  $\pm$     2  &   -0.84  $\pm$  0.03  &   -0.94  $\pm$  0.02  &   -4.42  $\pm$  0.78  &   -2.25  $\pm$  0.03  &    5.1  $\pm$   0.4  &   586  /   469  \\  
GRB100918863  &     \band   & 562  $\pm$     3  &   -0.80 $\pm$  0.01  &   -0.84  $\pm$  0.00  &   -2.74  $\pm$  0.12  &   -2.74  $\pm$  0.12  &    205  $\pm$     1  &   709  /   352  \\                       &     \issm   & 612  $\pm$    10  &   -0.76  $\pm$  0.01  &   -0.79  $\pm$  0.01  &   -5.05  $\pm$  0.14  &   -2.37  $\pm$  0.02  &   18.9  $\pm$   0.2  &   714  /   352  \\  
GRB101014175  &     \band   & 210  $\pm$     4  &   -1.17 $\pm$  0.01  &   -1.22  $\pm$  0.01  &   -2.79  $\pm$  0.11  &   -2.79  $\pm$  0.11  &    625  $\pm$    12  &   356  /   237  \\                       &     \issm   & 218  $\pm$     5  &   -1.16  $\pm$  0.01  &   -1.20  $\pm$  0.01  &   -9.04  $\pm$  3.19  &   -2.61  $\pm$  0.02  &   14.6  $\pm$   0.4  &   365  /   237  \\  
GRB101023951  &     \band   & 185  $\pm$     7  &   -1.22 $\pm$  0.03  &   -1.28  $\pm$  0.02  &   -2.58  $\pm$  0.14  &   -2.58  $\pm$  0.14  &    220  $\pm$    10  &   1653  /   353  \\                       &     \issm   & 187  $\pm$     6  &   -1.12  $\pm$  0.03  &   -1.19  $\pm$  0.02  &   -3.61  $\pm$  0.48  &   -2.33  $\pm$  0.04  &    4.7  $\pm$   0.2  &   1649  /   353  \\  
GRB101126198  &     \band   & 135  $\pm$     1  &   -1.29 $\pm$  0.01  &   -1.37  $\pm$  0.00  &   -2.65  $\pm$  0.17  &   -2.65  $\pm$  0.17  &    211  $\pm$     1  &   890  /   470  \\                       &     \issm   & 140  $\pm$     1  &   -1.28  $\pm$  0.01  &   -1.34  $\pm$  0.01  &   -8.81  $\pm$  0.29  &   -2.51  $\pm$  0.01  &    2.5  $\pm$   0.1  &   893  /   470  \\  
GRB101231067  &     \band   & 214  $\pm$     2  &   -0.75 $\pm$  0.02  &   -0.83  $\pm$  0.01  &   -9.99  $\pm$  4.99  &   -9.99  $\pm$  4.99  &    251  $\pm$     3  &   531  /   353  \\                       &     \issm   & 216  $\pm$     4  &   -0.70  $\pm$  0.04  &   -0.78  $\pm$  0.03  &   -8.07  $\pm$  2.97  &   -5.19  $\pm$  0.07  &    4.6  $\pm$   0.2  &   531  /   353  \\  
GRB110301214  &     \band   & 110  $\pm$     1  &   -0.83 $\pm$  0.02  &   -0.99  $\pm$  0.02  &   -2.73  $\pm$  0.05  &   -2.73  $\pm$  0.05  &   4242  $\pm$   124  &   713  /   470  \\                       &     \issm   & 110  $\pm$     2  &   -0.59  $\pm$  0.04  &   -0.80  $\pm$  0.03  &   -4.03  $\pm$  0.18  &   -2.42  $\pm$  0.01  &   22.0  $\pm$   0.9  &   690  /   470  \\  
GRB110622158  &     \band   & 105  $\pm$     1  &   -0.64 $\pm$  0.03  &   -0.83  $\pm$  0.02  &   -2.44  $\pm$  0.04  &   -2.44  $\pm$  0.04  &    541  $\pm$    25  &   1973  /   471  \\                       &     \issm   & 114  $\pm$     2  &   -0.17  $\pm$  0.06  &   -0.52  $\pm$  0.03  &   -3.28  $\pm$  0.14  &   -2.15  $\pm$  0.01  &    2.9  $\pm$   0.1  &   1997  /   471  \\  
GRB110625881  &     \band   & 179  $\pm$     4  &   -0.77 $\pm$  0.02  &   -0.87  $\pm$  0.01  &   -2.33  $\pm$  0.04  &   -2.33  $\pm$  0.04  &    929  $\pm$    26  &   1285  /   470  \\                       &     \issm   & 210  $\pm$     3  &   -0.53  $\pm$  0.05  &   -0.68  $\pm$  0.03  &   -3.16  $\pm$  0.14  &   -2.05  $\pm$  0.01  &   17.4  $\pm$   0.3  &   1250  /   470  \\  
GRB110717319  &     \band   & 376  $\pm$     5  &   -1.01 $\pm$  0.01  &   -1.05  $\pm$  0.01  &   -9.37  $\pm$  1.57  &   -9.37  $\pm$  1.57  &     98  $\pm$     1  &   813  /   470  \\                       &     \issm   & 370  $\pm$     7  &   -0.98  $\pm$  0.01  &   -1.01  $\pm$  0.01  &   -10.00  $\pm$  1.50  &   -5.69  $\pm$  0.02  &    5.1  $\pm$   0.1  &   813  /   470  \\  
GRB110729142  &     \band   & 307  $\pm$    11  &   -1.02 $\pm$  0.02  &   -1.07  $\pm$  0.01  &   -2.21  $\pm$  0.15  &   -2.21  $\pm$  0.15  &     35  $\pm$     1  &   838  /   473  \\                       &     \issm   & 390  $\pm$    26  &   -0.91  $\pm$  0.07  &   -0.97  $\pm$  0.06  &   -2.89  $\pm$  0.31  &   -1.98  $\pm$  0.03  &    1.6  $\pm$   0.0  &   835  /   473  \\  
GRB110731465  &     \band   & 307  $\pm$    15  &   -0.87 $\pm$  0.02  &   -0.92  $\pm$  0.02  &   -2.88  $\pm$  0.65  &   -2.88  $\pm$  0.65  &    565  $\pm$    14  &   423  /   354  \\                       &     \issm   & 322  $\pm$     9  &   -0.86  $\pm$  0.02  &   -0.90  $\pm$  0.02  &   -10.00  $\pm$  1.50  &   -2.64  $\pm$  0.04  &   23.3  $\pm$   0.6  &   427  /   354  \\  
GRB110825102  &     \band   & 262  $\pm$     2  &   -1.07 $\pm$  0.01  &   -1.12  $\pm$  0.01  &   -2.72  $\pm$  0.31  &   -2.72  $\pm$  0.31  &    177  $\pm$     1  &   697  /   473  \\                       &     \issm   & 267  $\pm$     1  &   -1.05  $\pm$  0.01  &   -1.09  $\pm$  0.01  &   -10.00  $\pm$  1.50  &   -2.58  $\pm$  0.02  &    5.6  $\pm$   0.0  &   698  /   473  \\  
GRB110921912  &     \band   & 513  $\pm$    20  &   -0.88 $\pm$  0.01  &   -0.91  $\pm$  0.01  &   -2.36  $\pm$  0.09  &   -2.36  $\pm$  0.09  &    283  $\pm$     5  &   506  /   356  \\                       &     \issm   & 678  $\pm$    43  &   -0.78  $\pm$  0.04  &   -0.82  $\pm$  0.03  &   -2.89  $\pm$  0.19  &   -2.00  $\pm$  0.03  &   22.4  $\pm$   0.4  &   489  /   356  \\  
GRB111003465  &     \band   & 205  $\pm$     7  &   -0.95 $\pm$  0.02  &   -1.02  $\pm$  0.02  &   -2.43  $\pm$  0.10  &   -2.43  $\pm$  0.10  &    394  $\pm$    16  &   627  /   473  \\                       &     \issm   & 228  $\pm$     9  &   -0.86  $\pm$  0.06  &   -0.94  $\pm$  0.04  &   -3.76  $\pm$  0.57  &   -2.16  $\pm$  0.04  &    9.3  $\pm$   0.4  &   625  /   473  \\  
GRB111216389  &     \band   & 165  $\pm$     5  &   -0.91 $\pm$  0.03  &   -1.00  $\pm$  0.03  &   -2.30  $\pm$  0.06  &   -2.30  $\pm$  0.06  &    199  $\pm$     9  &   734  /   352  \\                       &     \issm   & 197  $\pm$     7  &   -0.79  $\pm$  0.07  &   -0.90  $\pm$  0.05  &   -3.30  $\pm$  0.30  &   -2.04  $\pm$  0.02  &    3.5  $\pm$   0.2  &   730  /   352  \\  
GRB111220486  &     \band   & 300  $\pm$    10  &   -1.05 $\pm$  0.01  &   -1.09  $\pm$  0.01  &   -2.30  $\pm$  0.07  &   -2.30  $\pm$  0.07  &    308  $\pm$     6  &   474  /   353  \\                       &     \issm   & 371  $\pm$    15  &   -0.96  $\pm$  0.01  &   -1.02  $\pm$  0.00  &   -3.01  $\pm$  0.15  &   -2.03  $\pm$  0.02  &   13.2  $\pm$   0.2  &   467  /   353  \\  
GRB120119170  &     \band   & 208  $\pm$     1  &   -1.03 $\pm$  0.01  &   -1.09  $\pm$  0.01  &   -2.54  $\pm$  0.10  &   -2.54  $\pm$  0.10  &    207  $\pm$     1  &   774  /   469  \\                       &     \issm   & 226  $\pm$     3  &   -0.97  $\pm$  0.01  &   -1.04  $\pm$  0.01  &   -4.37  $\pm$  0.30  &   -2.27  $\pm$  0.02  &    5.0  $\pm$   0.0  &   773  /   469  \\  
GRB120129580  &     \band   & 299  $\pm$     7  &   -0.68 $\pm$  0.02  &   -0.74  $\pm$  0.02  &   -2.56  $\pm$  0.07  &   -2.56  $\pm$  0.07  &   3845  $\pm$   100  &   392  /   236  \\                       &     \issm   & 337  $\pm$     8  &   -0.47  $\pm$  0.03  &   -0.56  $\pm$  0.03  &   -3.28  $\pm$  0.18  &   -2.16  $\pm$  0.02  &   157.3  $\pm$   2.4  &   346  /   236  \\  
GRB120204054  &     \band   & 163  $\pm$     2  &   -1.08 $\pm$  0.01  &   -1.16  $\pm$  0.01  &   -2.58  $\pm$  0.05  &   -2.58  $\pm$  0.05  &    612  $\pm$    11  &   1763  /   470  \\                       &     \issm   & 171  $\pm$     3  &   -1.01  $\pm$  0.02  &   -1.09  $\pm$  0.02  &   -4.31  $\pm$  0.30  &   -2.33  $\pm$  0.01  &    9.7  $\pm$   0.2  &   1760  /   470  \\  
GRB120226871  &     \band   & 301  $\pm$    11  &   -0.89 $\pm$  0.02  &   -0.94  $\pm$  0.02  &   -2.26  $\pm$  0.08  &   -2.26  $\pm$  0.08  &    231  $\pm$     8  &   1338  /   470  \\                       &     \issm   & 397  $\pm$    21  &   -0.76  $\pm$  0.04  &   -0.83  $\pm$  0.03  &   -2.89  $\pm$  0.22  &   -1.96  $\pm$  0.02  &   10.3  $\pm$   0.2  &   1318  /   470  \\  
GRB120328268  &     \band   & 194  $\pm$     4  &   -0.78 $\pm$  0.02  &   -0.87  $\pm$  0.01  &   -2.00  $\pm$  0.02  &   -2.00  $\pm$  0.02  &    799  $\pm$    21  &   1414  /   471  \\                       &     \issm   & 385  $\pm$    23  &   -0.47  $\pm$  0.05  &   -0.66  $\pm$  0.02  &   -2.35  $\pm$  0.05  &   -1.76  $\pm$  0.03  &   22.7  $\pm$   0.3  &   1357  /   471  \\  
GRB120426090  &     \band   & 135  $\pm$     3  &   -0.59 $\pm$  0.03  &   -0.74  $\pm$  0.02  &   -2.94  $\pm$  0.12  &   -2.94  $\pm$  0.12  &   4721  $\pm$   208  &   524  /   352  \\                       &     \issm   & 132  $\pm$     3  &   -0.28  $\pm$  0.07  &   -0.49  $\pm$  0.05  &   -4.49  $\pm$  0.42  &   -2.55  $\pm$  0.03  &   27.1  $\pm$   1.6  &   501  /   352  \\  
GRB120624933  &     \band   & 583  $\pm$    83  &   -0.97 $\pm$  0.05  &   -0.99  $\pm$  0.05  &   -2.05  $\pm$  0.16  &   -2.05  $\pm$  0.16  &     18  $\pm$     1  &   2055  /   469  \\                       &     \issm   &1107  $\pm$   457  &   -0.96  $\pm$  0.07  &   -0.98  $\pm$  0.07  &   -2.62  $\pm$  0.51  &   -1.76  $\pm$  0.10  &    1.6  $\pm$   0.1  &   2058  /   469  \\  
GRB120707800  &     \band   & 181  $\pm$    13  &   -1.08 $\pm$  0.03  &   -1.15  $\pm$  0.03  &   -2.37  $\pm$  0.05  &   -2.37  $\pm$  0.05  &    708  $\pm$    29  &   1173  /   352  \\                       &     \issm   & 189  $\pm$     6  &   -0.76  $\pm$  0.10  &   -0.91  $\pm$  0.19  &   -2.83  $\pm$  0.13  &   -2.14  $\pm$  0.06  &   15.2  $\pm$   0.3  &   1167  /   352  \\  
GRB120711115  &     \band   &1277  $\pm$    31  &   -0.95 $\pm$  0.01  &   -0.96  $\pm$  0.01  &   -3.11  $\pm$  0.13  &   -3.11  $\pm$  0.13  &    385  $\pm$     2  &   577  /   353  \\                       &     \issm   &1360  $\pm$    27  &   -0.95  $\pm$  0.01  &   -0.96  $\pm$  0.01  &   -8.68  $\pm$  1.64  &   -2.75  $\pm$  0.02  &   54.7  $\pm$   0.2  &   594  /   353  \\  
GRB130306991  &     \band   & 307  $\pm$    15  &   -0.75 $\pm$  0.03  &   -0.81  $\pm$  0.03  &   -2.62  $\pm$  0.11  &   -2.62  $\pm$  0.11  &    301  $\pm$     6  &   2204  /   470  \\                       &     \issm   & 323  $\pm$     5  &   -0.50  $\pm$  0.11  &   -0.59  $\pm$  0.11  &   -3.70  $\pm$  0.24  &   -2.28  $\pm$  0.11  &   12.6  $\pm$   0.2  &   2200  /   470  \\  
GRB130327350  &     \band   & 375  $\pm$     8  &   -0.61 $\pm$  0.02  &   -0.66  $\pm$  0.01  &   -9.37  $\pm$  2.22  &   -9.37  $\pm$  2.22  &    287  $\pm$     5  &   1057  /   470  \\                       &     \issm   & 379  $\pm$     8  &   -0.57  $\pm$  0.02  &   -0.61  $\pm$  0.02  &   -10.00  $\pm$  1.50  &   -5.54  $\pm$  0.03  &   16.7  $\pm$   0.3  &   1063  /   470  \\  
GRB130502327  &     \band   & 293  $\pm$     5  &   -0.50 $\pm$  0.01  &   -0.57  $\pm$  0.01  &   -2.36  $\pm$  0.04  &   -2.36  $\pm$  0.04  &    972  $\pm$    16  &   1361  /   473  \\                       &     \issm   & 354  $\pm$     5  &   -0.35  $\pm$  0.02  &   -0.44  $\pm$  0.02  &   -3.72  $\pm$  0.16  &   -2.02  $\pm$  0.01  &   41.6  $\pm$   0.4  &   1338  /   473  \\  
GRB130504978  &     \band   & 654  $\pm$    29  &   -1.20 $\pm$  0.01  &   -1.21  $\pm$  0.01  &   -2.27  $\pm$  0.07  &   -2.27  $\pm$  0.07  &    232  $\pm$     2  &   2120  /   470  \\                       &     \issm   & 867  $\pm$    41  &   -1.18  $\pm$  0.01  &   -1.19  $\pm$  0.01  &   -3.05  $\pm$  0.17  &   -2.00  $\pm$  0.01  &   18.1  $\pm$   0.2  &   2125  /   470  \\  
GRB130518580  &     \band   & 387  $\pm$    10  &   -0.87 $\pm$  0.01  &   -0.91  $\pm$  0.01  &   -2.22  $\pm$  0.05  &   -2.22  $\pm$  0.05  &    330  $\pm$     6  &   768  /   354  \\                       &     \issm   & 539  $\pm$    20  &   -0.78  $\pm$  0.02  &   -0.83  $\pm$  0.02  &   -2.93  $\pm$  0.14  &   -1.92  $\pm$  0.02  &   20.3  $\pm$   0.2  &   769  /   354  \\  
GRB130606497  &     \band   & 515  $\pm$    21  &   -1.13 $\pm$  0.01  &   -1.15  $\pm$  0.01  &   -2.10  $\pm$  0.02  &   -2.10  $\pm$  0.02  &    544  $\pm$     6  &   892  /   236  \\                       &     \issm   & 926  $\pm$    43  &   -1.03  $\pm$  0.01  &   -1.07  $\pm$  0.01  &   -2.35  $\pm$  0.04  &   -1.86  $\pm$  0.01  &   37.5  $\pm$   0.3  &   919  /   236  \\  
GRB130609902  &     \band   & 531  $\pm$    13  &   -0.98 $\pm$  0.02  &   -1.01  $\pm$  0.02  &   -9.37  $\pm$  1.77  &   -9.37  $\pm$  1.77  &     47  $\pm$     1  &   822  /   354  \\                       &     \issm   & 539  $\pm$    31  &   -0.96  $\pm$  0.02  &   -0.98  $\pm$  0.01  &   -9.37  $\pm$  3.80  &   -5.45  $\pm$  0.03  &    3.7  $\pm$   0.1  &   822  /   354  \\  
GRB130720582  &     \band   &  65  $\pm$     3  &   -0.95 $\pm$  0.03  &   -1.18  $\pm$  0.03  &   -2.39  $\pm$  0.02  &   -2.39  $\pm$  0.02  &    451  $\pm$    24  &   2608  /   469  \\                       &     \issm   &  66  $\pm$     1  &   -0.19  $\pm$  0.05  &   -0.83  $\pm$  0.06  &   -2.86  $\pm$  0.03  &   -2.16  $\pm$  0.01  &    1.7  $\pm$   0.0  &   2570  /   469  \\  
GRB131028076  &     \band   & 848  $\pm$    15  &   -0.64 $\pm$  0.01  &   -0.66  $\pm$  0.01  &   -2.55  $\pm$  0.03  &   -2.55  $\pm$  0.03  &    791  $\pm$     6  &   1132  /   353  \\                       &     \issm   & 952  $\pm$     9  &   -0.61  $\pm$  0.00  &   -0.63  $\pm$  0.00  &   -6.16  $\pm$  0.22  &   -2.24  $\pm$  0.01  &   125.0  $\pm$   0.5  &   2119  /   353  \\  
GRB131118958  &     \band   & 332  $\pm$    14  &   -0.69 $\pm$  0.02  &   -0.75  $\pm$  0.02  &   -9.37  $\pm$  1.67  &   -9.37  $\pm$  1.67  &    195  $\pm$     4  &   1105  /   237  \\                       &     \issm   & 313  $\pm$     9  &   -0.39  $\pm$  0.13  &   -0.47  $\pm$  0.26  &   -4.43  $\pm$  0.87  &   -3.72  $\pm$  0.26  &    8.6  $\pm$   0.2  &   1092  /   237  \\  
GRB131231198  &     \band   & 218  $\pm$     6  &   -1.20 $\pm$  0.01  &   -1.25  $\pm$  0.01  &   -2.41  $\pm$  0.04  &   -2.41  $\pm$  0.04  &   1119  $\pm$    18  &   1350  /   355  \\                       &     \issm   & 232  $\pm$     4  &   -1.08  $\pm$  0.02  &   -1.15  $\pm$  0.05  &   -3.10  $\pm$  0.13  &   -2.17  $\pm$  0.05  &   31.5  $\pm$   0.3  &   1315  /   355  \\  
GRB140306146  &     \band   &1529  $\pm$    73  &   -1.01 $\pm$  0.01  &   -1.02  $\pm$  0.01  &   -5.09  $\pm$  1.80  &   -5.09  $\pm$  1.80  &    126  $\pm$     1  &   1492  /   355  \\                       &     \issm   &1535  $\pm$    62  &   -1.00  $\pm$  0.01  &   -1.01  $\pm$  0.01  &   -10.00  $\pm$  1.50  &   -4.05  $\pm$  0.02  &   17.7  $\pm$   0.2  &   1495  /   355  \\  
GRB140416060  &     \band   &  97  $\pm$     3  &   -1.15 $\pm$  0.01  &   -1.27  $\pm$  0.01  &   -2.37  $\pm$  0.03  &   -2.37  $\pm$  0.03  &   1056  $\pm$    75  &   2323  /   237  \\                       &     \issm   & 101  $\pm$     3  &   -0.86  $\pm$  0.06  &   -1.08  $\pm$  0.12  &   -2.93  $\pm$  0.11  &   -2.16  $\pm$  0.06  &    9.8  $\pm$   0.2  &   2315  /   237  \\  
GRB140508128  &     \band   & 264  $\pm$    13  &   -1.01 $\pm$  0.02  &   -1.07  $\pm$  0.02  &   -2.11  $\pm$  0.04  &   -2.11  $\pm$  0.04  &    312  $\pm$    11  &   1153  /   235  \\                       &     \issm   & 434  $\pm$    44  &   -0.83  $\pm$  0.06  &   -0.93  $\pm$  0.04  &   -2.41  $\pm$  0.10  &   -1.87  $\pm$  0.02  &   12.4  $\pm$   0.2  &   1164  /   235  \\  
GRB140523129  &     \band   & 269  $\pm$     7  &   -0.90 $\pm$  0.01  &   -0.96  $\pm$  0.01  &   -2.69  $\pm$  0.13  &   -2.69  $\pm$  0.13  &    632  $\pm$     9  &   765  /   471  \\                       &     \issm   & 285  $\pm$     4  &   -0.83  $\pm$  0.01  &   -0.89  $\pm$  0.01  &   -4.71  $\pm$  0.37  &   -2.38  $\pm$  0.01  &   21.4  $\pm$   0.3  &   760  /   471  \\  
GRB140810782  &     \band   & 309  $\pm$     6  &   -0.88 $\pm$  0.01  &   -0.93  $\pm$  0.01  &   -2.41  $\pm$  0.06  &   -2.41  $\pm$  0.06  &    286  $\pm$     5  &   896  /   353  \\                       &     \issm   & 368  $\pm$    14  &   -0.75  $\pm$  0.03  &   -0.81  $\pm$  0.03  &   -3.17  $\pm$  0.20  &   -2.08  $\pm$  0.03  &   12.6  $\pm$   0.2  &   871  /   353  \\  
GRB150118409  &     \band   & 763  $\pm$    17  &   -0.84 $\pm$  0.01  &   -0.86  $\pm$  0.01  &   -3.51  $\pm$  0.25  &   -3.51  $\pm$  0.25  &    332  $\pm$     3  &   2545  /   469  \\                       &     \issm   & 795  $\pm$    18  &   -0.83  $\pm$  0.01  &   -0.84  $\pm$  0.01  &   -10.00  $\pm$  1.50  &   -3.07  $\pm$  0.02  &   39.9  $\pm$   0.3  &   2558  /   469  \\  
GRB150330828  &     \band   & 265  $\pm$     5  &   -1.01 $\pm$  0.01  &   -1.06  $\pm$  0.01  &   -2.25  $\pm$  0.04  &   -2.25  $\pm$  0.04  &    202  $\pm$     3  &   1708  /   469  \\                       &     \issm   & 346  $\pm$    11  &   -0.90  $\pm$  0.02  &   -0.97  $\pm$  0.02  &   -2.86  $\pm$  0.13  &   -1.98  $\pm$  0.01  &    7.7  $\pm$   0.1  &   1683  /   469  \\  
GRB150403913  &     \band   & 402  $\pm$    16  &   -0.82 $\pm$  0.02  &   -0.86  $\pm$  0.02  &   -2.09  $\pm$  0.04  &   -2.09  $\pm$  0.04  &    437  $\pm$    10  &   624  /   355  \\                       &     \issm   & 721  $\pm$    45  &   -0.67  $\pm$  0.03  &   -0.74  $\pm$  0.02  &   -2.49  $\pm$  0.07  &   -1.80  $\pm$  0.02  &   29.0  $\pm$   0.4  &   578  /   355  \\  
GRB150627183  &     \band   & 243  $\pm$     5  &   -0.92 $\pm$  0.01  &   -0.98  $\pm$  0.01  &   -2.19  $\pm$  0.02  &   -2.19  $\pm$  0.02  &    664  $\pm$    11  &   1109  /   355  \\                       &     \issm   & 334  $\pm$     8  &   -0.76  $\pm$  0.02  &   -0.86  $\pm$  0.01  &   -2.73  $\pm$  0.07  &   -1.93  $\pm$  0.01  &   23.0  $\pm$   0.2  &   1057  /   355  \\  
GRB150902733  &     \band   & 368  $\pm$     7  &   -0.49 $\pm$  0.01  &   -0.55  $\pm$  0.01  &   -2.35  $\pm$  0.04  &   -2.35  $\pm$  0.04  &   1085  $\pm$    17  &   761  /   470  \\                       &     \issm   & 472  $\pm$     7  &   -0.30  $\pm$  0.03  &   -0.38  $\pm$  0.02  &   -3.24  $\pm$  0.10  &   -1.97  $\pm$  0.01  &   68.3  $\pm$   0.6  &   656  /   470  \\  
GRB160802259  &     \band   & 295  $\pm$     5  &   -0.54 $\pm$  0.02  &   -0.62  $\pm$  0.02  &   -2.47  $\pm$  0.07  &   -2.47  $\pm$  0.07  &    863  $\pm$    20  &   314  /   237  \\                       &     \issm   & 346  $\pm$     9  &   -0.40  $\pm$  0.01  &   -0.49  $\pm$  0.01  &   -3.73  $\pm$  0.13  &   -2.10  $\pm$  0.02  &   36.2  $\pm$   0.7  &   298  /   237  \\  
GRB160905471  &     \band   &1063  $\pm$    52  &   -0.89 $\pm$  0.01  &   -0.90  $\pm$  0.01  &   -3.01  $\pm$  0.27  &   -3.01  $\pm$  0.27  &    237  $\pm$     2  &   730  /   356  \\                       &     \issm   &1161  $\pm$    20  &   -0.89  $\pm$  0.01  &   -0.90  $\pm$  0.01  &   -10.00  $\pm$  0.00  &   -2.66  $\pm$  0.02  &   33.5  $\pm$   0.2  &   736  /   356  \\  
GRB160910722  &     \band   & 335  $\pm$     7  &   -0.76 $\pm$  0.01  &   -0.82  $\pm$  0.01  &   -2.23  $\pm$  0.03  &   -2.23  $\pm$  0.03  &    632  $\pm$    11  &   786  /   469  \\                       &     \issm   & 460  $\pm$    11  &   -0.60  $\pm$  0.02  &   -0.67  $\pm$  0.04  &   -2.85  $\pm$  0.08  &   -1.92  $\pm$  0.02  &   33.1  $\pm$   0.3  &   746  /   469  \\
\hline
\label{tab:catalog:Band_IAP}
\end{longtable}
\normalsize
\end{appendix}

%
%

\bibliographystyle{aa} 
\bibliography{biblio.bib} 

\begin{thebibliography}{51}
\expandafter\ifx\csname natexlab\endcsname\relax\def\natexlab#1{#1}\fi

\bibitem[{{Abbott} {et~al.}(2017){Abbott}, {Abbott}, {Abbott}, {Acernese},
  {Ackley}, {Adams}, {Adams}, {Addesso}, {Adhikari}, {Adya}, \&
  et~al.}]{abbott}
{Abbott}, B.~P., {Abbott}, R., {Abbott}, T.~D., {et~al.} 2017, \apjl, 848, L13

\bibitem[{{Abdo} {et~al.}(2009){Abdo}, {Ackermann}, {Ajello}, {Asano}, \&
  {Atwood}}]{090902B2009}
{Abdo}, A.~A., {Ackermann}, M., {Ajello}, M., {Asano}, K., \& {Atwood}, W.~B.,
  e.~a. 2009, APJL, 706, L138

\bibitem[{{Ackermann} {et~al.}(2011){Ackermann}, {Ajello}, {Asano}, {Axelsson},
  \& {Baldini}}]{Ackermann2011ApJ...729..114A}
{Ackermann}, M., {Ajello}, M., {Asano}, K., {Axelsson}, M., \& {Baldini}, L.,
  e.~a. 2011, APJ, 729, 114

\bibitem[{{Ackermann} {et~al.}(2013){Ackermann}, {Ajello}, {Asano}, {Axelsson},
  {Baldini}, \& {Ballet}}]{LATCatalog2013}
{Ackermann}, M., {Ajello}, M., {Asano}, K., {et~al.} 2013, APJS, 209, 11

\bibitem[{{Ajello} {et~al.}(2019){Ajello}, {Arimoto}, {Axelsson}, {Baldini},
  {Barbiellini}, {Bastieri}, {Bellazzini}, {Bhat}, {Bissaldi}, \&
  {Blandford}}]{LatGrbCatalog2}
{Ajello}, M., {Arimoto}, M., {Axelsson}, M., {et~al.} 2019, \apj, 878, 52

\bibitem[{{Atwood} {et~al.}(2009){Atwood}, {Abdo}, {Ackermann}, {Althouse},
  {Anderson}, {Axelsson}, {Baldini}, {Ballet}, {Band}, {Barbiellini}, \&
  et~al.}]{Atwood2009ApJ6971071A}
{Atwood}, W.~B., {Abdo}, A.~A., {Ackermann}, M., {et~al.} 2009, \apj, 697, 1071

\bibitem[{{Atwood} {et~al.}(2013){Atwood}, {Baldini}, {Bregeon}, {Bruel},
  {Chekhtman}, {Cohen-Tanugi}, {Drlica-Wagner}, {Granot}, {Longo}, \&
  {Omodei}}]{Atwood2013}
{Atwood}, W.~B., {Baldini}, L., {Bregeon}, J., {et~al.} 2013, \apj, 774, 76

\bibitem[{{Axelsson} {et~al.}(2012){Axelsson}, {Baldini}, {Barbiellini},
  {Baring}, \& {Bellazzini}}]{Axelsson2012}
{Axelsson}, M., {Baldini}, L., {Barbiellini}, G., {Baring}, M.~G., \&
  {Bellazzini}, R., e.~a. 2012, APJL, 757, L31

\bibitem[{{Axelsson} \& {Borgonovo}(2015)}]{Axelsson2015}
{Axelsson}, M. \& {Borgonovo}, L. 2015, \mnras, 447, 3150

\bibitem[{{Band} {et~al.}(1993){Band}, {Matteson}, {Ford}, {Schaefer}, \&
  {Palmer}}]{Band1993}
{Band}, D., {Matteson}, J., {Ford}, L., {Schaefer}, B., \& {Palmer}, D., e.~a.
  1993, APJ, 413, 281

\bibitem[{{Beloborodov} \& {M{\'e}sz{\'a}ros}(2017)}]{BM17}
{Beloborodov}, A.~M. \& {M{\'e}sz{\'a}ros}, P. 2017, \ssr, 207, 87

\bibitem[{{Beniamini} \& {Granot}(2016)}]{BG16}
{Beniamini}, P. \& {Granot}, J. 2016, \mnras, 459, 3635

\bibitem[{{Beniamini} \& {Piran}(2013)}]{Beniamini2013}
{Beniamini}, P. \& {Piran}, T. 2013, \apj, 769, 69

\bibitem[{{Bernardini} {et~al.}(2017){Bernardini}, {Xie}, {Sizun}, {Piron},
  {Dong}, {Atteia}, {Antier}, {Daigne}, {Godet}, {Cordier}, \&
  {Wei}}]{bernardini2017}
{Bernardini}, M.~G., {Xie}, F., {Sizun}, P., {et~al.} 2017, Experimental
  Astronomy, 44, 113

\bibitem[{{Bloom} {et~al.}(2002){Bloom}, {Kulkarni}, {Price}, {Reichart},
  {Galama}, {Schmidt}, {Frail}, {Berger}, {McCarthy}, {Chevalier}, {Wheeler},
  {Halpern}, {Fox}, {Djorgovski}, {Harrison}, {Sari}, {Axelrod}, {Kimble},
  {Holtzman}, {Hurley}, {Frontera}, {Piro}, \& {Costa}}]{bloom}
{Bloom}, J.~S., {Kulkarni}, S.~R., {Price}, P.~A., {et~al.} 2002, \apjl, 572,
  L45

\bibitem[{{Bo{\v s}njak} \& {Daigne}(2014)}]{BD14}
{Bo{\v s}njak}, {\v Z}. \& {Daigne}, F. 2014, \aap, 568, A45

\bibitem[{{Bo{\v s}njak} {et~al.}(2009){Bo{\v s}njak}, {Daigne}, \&
  {Dubus}}]{BD09}
{Bo{\v s}njak}, {\v Z}., {Daigne}, F., \& {Dubus}, G. 2009, \aap, 498, 677

\bibitem[{{Burgess}(2019)}]{Burgess2019}
{Burgess}, J.~M. 2019, \aap, 629, A69

\bibitem[{{Burgess} {et~al.}(2015){Burgess}, {Ryde}, \& {Yu}}]{Burgess2015}
{Burgess}, J.~M., {Ryde}, F., \& {Yu}, H.-F. 2015, \mnras, 451, 1511

\bibitem[{{Crider} {et~al.}(1997){Crider}, {Liang}, {Smith}, {Preece},
  {Briggs}, {Pendleton}, {Paciesas}, {Band }, \& {Matteson}}]{Crider1997}
{Crider}, A., {Liang}, E.~P., {Smith}, I.~A., {et~al.} 1997, \apjl, 479, L39

\bibitem[{{Daigne} {et~al.}(2011){Daigne}, {Bo{\v s}njak}, \& {Dubus}}]{DB11}
{Daigne}, F., {Bo{\v s}njak}, {\v Z}., \& {Dubus}, G. 2011, \aap, 526, A110

\bibitem[{{Daigne} \& {Mochkovitch}(1998)}]{DM98}
{Daigne}, F. \& {Mochkovitch}, R. 1998, \mnras, 296, 275

\bibitem[{{D'Avanzo}(2015)}]{DAvanzo2015JHEAp7-73}
{D'Avanzo}, P. 2015, Journal of High Energy Astrophysics, 7, 73

\bibitem[{{Derishev}(2007)}]{derishev07}
{Derishev}, E.~V. 2007, \apss, 309, 157

\bibitem[{{Gehrels} {et~al.}(2005){Gehrels}, {Sarazin}, {O'Brien}, {Zhang},
  {Barbier}, {Barthelmy}, {Blustin}, {Burrows}, {Cannizzo}, {Cummings}, {Goad},
  {Holland}, {Hurkett}, {Kennea}, {Levan}, {Markwardt}, {Mason}, {Meszaros},
  {Page}, {Palmer}, {Rol}, {Sakamoto}, {Willingale}, {Angelini}, {Beardmore},
  {Boyd}, {Breeveld}, {Campana}, {Chester}, {Chincarini}, {Cominsky},
  {Cusumano}, {de Pasquale}, {Fenimore}, {Giommi}, {Gronwall}, {Grupe}, {Hill},
  {Hinshaw}, {Hjorth}, {Hullinger}, {Hurley}, {Klose}, {Kobayashi},
  {Kouveliotou}, {Krimm}, {Mangano}, {Marshall}, {McGowan}, {Moretti},
  {Mushotzky}, {Nakazawa}, {Norris}, {Nousek}, {Osborne}, {Page}, {Parsons},
  {Patel}, {Perri}, {Poole}, {Romano}, {Roming}, {Rosen}, {Sato}, {Schady},
  {Smale}, {Sollerman}, {Starling}, {Still}, {Suzuki}, {Tagliaferri},
  {Takahashi}, {Tashiro}, {Tueller}, {Wells}, {White}, \& {Wijers}}]{gehrels}
{Gehrels}, N., {Sarazin}, C.~L., {O'Brien}, P.~T., {et~al.} 2005, \nat, 437,
  851

\bibitem[{{Ghisellini} {et~al.}(2000){Ghisellini}, {Celotti}, \&
  {Lazzati}}]{Ghisellini2000}
{Ghisellini}, G., {Celotti}, A., \& {Lazzati}, D. 2000, \mnras, 313, L1

\bibitem[{{Giannios}(2008)}]{giannios}
{Giannios}, D. 2008, \aap, 480, 305

\bibitem[{{Gruber} {et~al.}(2014){Gruber}, {Goldstein}, {Weller von Ahlefeld},
  {Narayana Bhat}, \& {Bissaldi}}]{Gruber2014}
{Gruber}, D., {Goldstein}, A., {Weller von Ahlefeld}, V., {Narayana Bhat}, P.,
  \& {Bissaldi}, E., e.~a. 2014, APJS, 211, 12

\bibitem[{{Guiriec} {et~al.}(2011){Guiriec}, {Connaughton}, {Briggs},
  {Burgess}, \& {Ryde}}]{Guiriec2011}
{Guiriec}, S., {Connaughton}, V., {Briggs}, M.~S., {Burgess}, M., \& {Ryde},
  F., e.~a. 2011, APJL, 727, L33

\bibitem[{{Guiriec} {et~al.}(2015){Guiriec}, {Kouveliotou}, {Daigne}, {Zhang},
  \& {Hasco{\"e}t}}]{Guiriec2015ApJ...807..148G}
{Guiriec}, S., {Kouveliotou}, C., {Daigne}, F., {Zhang}, B., \& {Hasco{\"e}t},
  R., e.~a. 2015, APJ, 807, 148

\bibitem[{{Hjorth} {et~al.}(2005){Hjorth}, {Sollerman}, {Gorosabel}, {Granot},
  {Klose}, {Kouveliotou}, {Melinder}, {Ramirez-Ruiz}, {Starling}, {Thomsen},
  {Andersen}, {Fynbo}, {Jensen}, {Vreeswijk}, {Castro Cer{\'o}n}, {Jakobsson},
  {Levan}, {Pedersen}, {Rhoads}, {Tanvir}, {Watson}, \& {Wijers}}]{hjorth2}
{Hjorth}, J., {Sollerman}, J., {Gorosabel}, J., {et~al.} 2005, \apjl, 630, L117

\bibitem[{{Hjorth} {et~al.}(2003){Hjorth}, {Sollerman}, {M{\o}ller}, {Fynbo},
  {Woosley}, {Kouveliotou}, {Tanvir}, {Greiner}, {Andersen}, {Castro-Tirado},
  {Castro Cer{\'o}n}, {Fruchter}, {Gorosabel}, {Jakobsson}, {Kaper}, {Klose},
  {Masetti}, {Pedersen}, {Pedersen}, {Pian}, {Palazzi}, {Rhoads}, {Rol}, {van
  den Heuvel}, {Vreeswijk}, {Watson}, \& {Wijers}}]{hjorth1}
{Hjorth}, J., {Sollerman}, J., {M{\o}ller}, P., {et~al.} 2003, \nat, 423, 847

\bibitem[{{Kawabata} {et~al.}(2003){Kawabata}, {Deng}, {Wang}, {Mazzali},
  {Nomoto}, {Maeda}, {Tominaga}, {Umeda}, {Iye}, {Kosugi}, {Ohyama}, {Sasaki},
  {H{\"o}flich}, {Wheeler}, {Jeffery}, {Aoki}, {Kashikawa}, {Takata}, {Kawai},
  {Sakamoto}, {Urata}, {Yoshida}, {Tamagawa}, {Torii}, {Aoki}, {Kobayashi},
  {Komiyama}, {Mizumoto}, {Noumaru}, {Ogasawara}, {Sekiguchi}, {Shirasaki},
  {Totani}, {Watanabe}, \& {Yamada}}]{kawabata}
{Kawabata}, K.~S., {Deng}, J., {Wang}, L., {et~al.} 2003, \apjl, 593, L19

\bibitem[{{Kobayashi} {et~al.}(1997){Kobayashi}, {Piran}, \&
  {Sari}}]{kobayashi}
{Kobayashi}, S., {Piran}, T., \& {Sari}, R. 1997, \apj, 490, 92

\bibitem[{{Massaro} {et~al.}(2010){Massaro}, {Grindlay}, \&
  {Paggi}}]{Massaro2010ApJ}
{Massaro}, F., {Grindlay}, J.~E., \& {Paggi}, A. 2010, APJL, 714, L299

\bibitem[{{McKinney} \& {Uzdensky}(2012)}]{mckinney}
{McKinney}, J.~C. \& {Uzdensky}, D.~A. 2012, \mnras, 419, 573

\bibitem[{{Nakar}(2007)}]{Nakar2007PhysRep442-166}
{Nakar}, E. 2007, \physrep, 442, 166

\bibitem[{{Narayana Bhat} {et~al.}(2016){Narayana Bhat}, {Meegan}, {von
  Kienlin}, {Paciesas}, {Briggs}, {Burgess}, {Burns}, {Chaplin}, {Cleveland},
  {Collazzi}, {Connaughton}, {Diekmann}, {Fitzpatrick}, {Gibby}, {Giles},
  {Goldstein}, {Greiner}, {Jenke}, {Kippen}, {Kouveliotou}, {Mailyan},
  {McBreen}, {Pelassa}, {Preece}, {Roberts}, {Sparke}, {Stanbro}, {Veres},
  {Wilson-Hodge}, {Xiong}, {Younes}, {Yu}, \& {Zhang}}]{2016ApJS22328N}
{Narayana Bhat}, P., {Meegan}, C.~A., {von Kienlin}, A., {et~al.} 2016, \apjs,
  223, 28

\bibitem[{Neyman \& Pearson(1928)}]{NeymanPearson}
Neyman, J. \& Pearson, E.~S. 1928, On the Use and Interpretation of Certain
  Test Criteria for Purposes of Statistical Inference, Part I (London:
  Cambridge University Press)

\bibitem[{{Pe'er} \& {Zhang}(2006)}]{peerzhang}
{Pe'er}, A. \& {Zhang}, B. 2006, \apj, 653, 454

\bibitem[{{Preece} {et~al.}(1998){Preece}, {Briggs}, {Mallozzi}, {Pendleton},
  \& {Paciesas}}]{Preece1998}
{Preece}, R.~D., {Briggs}, M.~S., {Mallozzi}, R.~S., {Pendleton}, G.~N., \&
  {Paciesas}, W.~S., e.~a. 1998, APJL, 506, L23

\bibitem[{{Rees} \& {Meszaros}(1994)}]{rees}
{Rees}, M.~J. \& {Meszaros}, P. 1994, \apjl, 430, L93

\bibitem[{{Sironi} {et~al.}(2015){Sironi}, {Petropoulou}, \&
  {Giannios}}]{sironi}
{Sironi}, L., {Petropoulou}, M., \& {Giannios}, D. 2015, \mnras, 450, 183

\bibitem[{{Stanek} {et~al.}(2003){Stanek}, {Matheson}, {Garnavich}, {Martini},
  {Berlind}, {Caldwell}, {Challis}, {Brown}, {Schild}, {Krisciunas}, {Calkins},
  {Lee}, {Hathi}, {Jansen}, {Windhorst}, {Echevarria}, {Eisenstein}, {Pindor},
  {Olszewski}, {Harding}, {Holland}, \& {Bersier}}]{stanek}
{Stanek}, K.~Z., {Matheson}, T., {Garnavich}, P.~M., {et~al.} 2003, \apjl, 591,
  L17

\bibitem[{{Tierney} {et~al.}(2013){Tierney}, {McBreen}, {Preece},
  {Fitzpatrick}, \& {Foley}}]{Tierney2013AA}
{Tierney}, D., {McBreen}, S., {Preece}, R.~D., {Fitzpatrick}, G., \& {Foley},
  S., e.~a. 2013, AAP, 550, A102

\bibitem[{Wei {et~al.}(2016)Wei, Cordier, Antier, Antilogus, Atteia, Bajat,
  Basa, Beckmann, Bernardini, Boissier, Bouchet, Burwitz, Claret, Dai, Daigne,
  Deng, Dornic, Feng, Foglizzo, Gao, Gehrels, Godet, Goldwurm, Gonzalez,
  Gosset, G{\"o}tz, Gouiffes, Grise, Gros, Guilet, Han, Huang, Huang, Jouret,
  Klotz, La~Marle, Lachaud, Le~Floch, Lee, Leroy, Li, Li, Li, Liang, Lyu,
  Mercier, Migliori, Mochkovitch, O'Brien, Osborne, Paul, Perinati, Petitjean,
  Piron, Qiu, Rau, Rodriguez, Schanne, Tanvir, Vangioni, Vergani, Wang, Wang,
  Wang, Wang, Watson, Webb, Wei, Willingale, Wu, Wu, Xin, Xu, Yu, Yu, Yu,
  Zhang, Zhang, Zhang, \& Zhou}]{wei2016}
Wei, J., Cordier, B., Antier, S., {et~al.} 2016, {The Deep and Transient
  Universe in the SVOM Era: New Challenges and Opportunities - Scientific
  prospects of the SVOM mission}, report on the Scientific prospects of the
  SVOM mission. Proceedings of the Workshop held from 11th to 15th April 2016
  at Les Houches School of Physics, France

\bibitem[{Wilks(1938)}]{wilks1938}
Wilks, S.~S. 1938, Ann. Math. Statist., 9, 60

\bibitem[{{Woosley} \& {Bloom}(2006)}]{Woosley2006ARAA44-507}
{Woosley}, S.~E. \& {Bloom}, J.~S. 2006, \araa, 44, 507

\bibitem[{{Yassine} {et~al.}(2017){Yassine}, {Piron}, {Mochkovitch}, \&
  {Daigne}}]{Yassine2017}
{Yassine}, M., {Piron}, F., {Mochkovitch}, R., \& {Daigne}, F. 2017, \aap, 606,
  A93

\bibitem[{{Yu} {et~al.}(2015){Yu}, {van Eerten}, {Greiner}, {Sari}, \&
  {Narayana Bhat}}]{Yu2015A&A}
{Yu}, H.-F., {van Eerten}, H.~J., {Greiner}, J., {Sari}, R., \& {Narayana
  Bhat}, P., e.~a. 2015, AAP, 583, A129

\bibitem[{{Zhang} {et~al.}(2016){Zhang}, {Uhm}, {Connaughton}, {Briggs}, \&
  {Zhang}}]{zhang2016}
{Zhang}, B.-B., {Uhm}, Z.~L., {Connaughton}, V., {Briggs}, M.~S., \& {Zhang},
  B. 2016, \apj, 816, 72

\end{thebibliography}
\end{document}